\documentclass[english,superscriptaddress,showpacs,showkeys]{revtex4-2}
\usepackage[T1]{fontenc}
\usepackage{tikz}
\usepackage[compat=1.1.0]{tikz-feynman}
\usepackage[utf8]{inputenc}
\usepackage{color}
\usepackage{babel}
\usepackage{array}
\usepackage{units}
\usepackage{bbold}
\usepackage{amsmath}
\usepackage{amssymb}
\usepackage{graphicx}
\usepackage{bm}
\usepackage{subcaption}
\usepackage{todonotes}
\usepackage[pdfusetitle,
bookmarks=true,bookmarksnumbered=false,bookmarksopen=true,bookmarksopenlevel=2,
breaklinks=true,pdfborder={0 0 0},pdfborderstyle={},backref=false,colorlinks=true]
{hyperref}
\hypersetup{
	citecolor=blue,linkcolor=blue,urlcolor=blue}



\begin{document}
	
	\title{Soft photon approximation in a laser field: applications}
	
	\author{P.A. Krachkov}\email{P.A.Krachkov@inp.nsk.su}
	\author{Simon.V. Sorokin}\email{S.Sorokin1@g.nsu.ru}
	\affiliation{Budker Institute of Nuclear Physics, SB RAS, Novosibirsk, 630090, Russia}
	\affiliation{Novosibirsk State University, 630090 Novosibirsk, Russia}
	
	\date{\today}
\begin{abstract}
	We apply the soft photon approximation to QED processes in a strong plane-wave laser field. New compact expressions for the emission factors are presented, which simplify the generalization to multiple soft photon emission. The method is applied to nonlinear single, double, and $N$-photon Compton scattering. For single Compton scattering, we show that the soft photon approximation reproduces the exact spectrum with accuracy $\mathcal{O}(\omega/\varepsilon)$, significantly outperforming the classical radiation theory. For double Compton scattering, we derive analytical expressions for the differential probability in two kinematic regimes: when both photons are soft, and when one photon is soft while the other has arbitrary energy. We present a compact analytical expression for the amplitude of $N$-photon emission within the soft photon approximation. A numerical implementation in Wolfram Mathematica for single and double Compton scattering is provided.
\end{abstract}
\maketitle

\section{Introduction}\label{sec:intr}
The development of high-intensity laser facilities has opened new frontiers in strong-field quantum electrodynamics (QED). Modern and planned experiments such as LUXE at DESY~\cite{Abramowicz2021}, E-320 at SLAC FACET-II~\cite{yakimenko2019facet}, and {ELI-NP} in Romania~\cite{gales2018extreme}  aim to probe the nonperturbative regime of QED, where the dimensionless intensity parameter $\xi = |e|E_0/(m\omega_0)$ approaches or exceeds unity. In this regime, the interaction of an electron with a laser field cannot be treated perturbatively. Accurate theoretical descriptions of QED processes in the presence of a strong plane-wave background are of  importance for both  interpretation of experimental data and planning of future measurements.

The theoretical description of basic strong-field QED processes like nonlinear Compton scattering  and nonlinear Breit-Wheeler pair production has been studied  in detail by approximating the laser field as a plane wave; see reviews \cite{Mitter1975,Ritus1985,Baier1998,Ehlotzky2009, DiPiazza2012, Fedotov2022a}. The structure of the results in a plane wave field is much more complex than in the vacuum case, even for $1\rightarrow2$ processes. The processes  $1\rightarrow3$ and $2\rightarrow2$, such as double nonlinear Compton scattering, electron trident production, electron-positron annihilation into two photons, have a much more complex structure compared to the $1\rightarrow2$ processes.

The complexity of the results increases exponentially with the number of particles in the initial and final states. Thus, developing approximate methods is critical. A powerful tool for simplifying such calculations is the soft photon approximation. In standard QED, this approximation allows one to express the amplitude for a process with an additional soft photon in terms of the amplitude of the corresponding process without this photon, with corrections of order $\omega/\varepsilon$, where $\omega$ is the photon energy and $\varepsilon$ is the characteristic energy scale of the hard process \cite{LL4}. This approximation is widely used in high-energy physics to account for radiative corrections and to regulate infrared divergences.

In the presence of a strong laser field, the application of the soft photon approximation is complicated by the fact that the electron wave functions are no longer plane waves but Volkov states, which depend nontrivially on the field phase. In our previous paper \cite{Krachkov24}, we developed a systematic soft photon approximation for QED processes in a plane-wave background. The key improvement over the standard classical radiation theory \cite{landau1975classical,Ritus1985,Baier1998} is the exact treatment of the phase factor: while classical approaches neglect the quantum recoil of the electron in the phase, our approximation retains the recoil effects exactly in the phase, approximating only the preexponential factors. This leads to a significantly better agreement with exact QED results, in regimes where classical radiation theory fails.

The main strength of the soft photon approximation lies in its ability to provide compact analytical expressions for processes with additional soft photon emission in terms of the matrix element of the corresponding process without this photon. This approach significantly simplifies the calculation of such processes and produces new important results for them. This approximation is particularly powerful for processes with multiple soft photon emission, where exact calculations are difficult to perform. By expressing the amplitude for $N$-photon emission through the amplitude for $(N-1)$-photon emission, the 
soft photon approximation provides a systematic and tractable framework for studying cascade processes in strong-field QED.

The main goal of the present work is to demonstrate the power and applicability of the soft photon approximation for the simplest QED processes in a strong laser field. We give a detailed review of the soft photon approximation and present new compact expressions for the emission factors that significantly simplify the generalization to $N$-photon emission. We also consider nonlinear single Compton scattering, nonlinear double Compton scattering, and nonlinear $N$-photon Compton scattering. For single Compton scattering, we perform a detailed comparison between the exact QED result, the soft photon approximation, and the classical radiation theory, establishing the domain of applicability of each method. For double Compton scattering, we derive simple analytical expressions for the differential probability in two kinematic regimes: when both photons are soft, and when one photon is soft while the other has arbitrary energy. We also present new compact analytical expressions for the amplitude and differential probability of $N$-photon emission within the soft photon approximation. This provides a systematic analytical description including interference effects between amplitudes with different emission orderings, going beyond the incoherent cascade treatment of Ref.~\cite{PhysRevD.99.096018}. Throughout the paper, we compare our analytical results with exact QED calculations where available, demonstrating that the soft photon approximation provides accurate results with corrections $\mathcal{O}(\omega/\varepsilon)$, significantly outperforming the classical approximation.

To provide a fast and adaptable numerical implementation of the developed formalism, we attach \texttt{Mathematica} code that allows the reader to reproduce all numerical results presented in the paper and to apply the method to other kinematic regimes and laser pulse shapes. We focus on the most experimentally relevant regime: an ultra-relativistic electron colliding head-on with a linearly polarized laser pulse. In this setup, the emitted photons are collimated within small angles around the electron momentum, which allows us to present compact formulas in the small-angle approximation. Nevertheless, the general expressions derived in the paper are valid for arbitrary collision geometries and can be easily implemented using the provided code.

The paper is organized as follows. In Sec.~\ref{sec:SPA}, we review the soft photon approximation developed in \cite{Krachkov24} and present new compact expressions for the emission factors. In Sec.~\ref{sec:NCS}, we apply the method to nonlinear Compton scattering, comparing the soft photon approximation with the classical and exact results. In Sec.~\ref{sec:DCS}, we consider nonlinear double Compton scattering and derive the probability for both kinematic regimes. In Sec.~\ref{sec:NR}, we describe the numerical implementation and discuss its computational performance. Finally, in Sec.~\ref{sec:concl}, we summarize our results and discuss possible extensions. Additional technical details are provided in the Appendix.

\section{Soft photon approximation}\label{sec:SPA}

Consider a generic QED process in the presence of a laser field with initial state $i$, final state $f$, and matrix element $S_0$. We also study the matrix element $S$ for the $i\to f\gamma$ process, which differs from the first one by an additional photon $\gamma$ in the final state. If the frequency of this photon is small in comparison with $\varepsilon_{char}$, the matrix element $S$ can be simply obtained from $S_0$ by the soft photon approximation. The main contribution to the matrix element $S$ comes from the region where the soft photon is emitted from the initial or final state electron/positron.  If $p$ and $k$ are the momenta of an external electron line and soft photon, the Green's function $G(p\pm k)$  is near the pole for small $\omega=k^0$. That is, when a photon is emitted from an initial or final electron, the emission process has a large formation length \cite{BAIER2005261,Ritus1985}. However, for photon emission from an internal electron line, the formation length is restricted by the hard subprocesses. 

As an example, we consider the matrix element $S_0$ for a process with one electron in the initial and  final states:
\begin{equation}\label{eq:M0}
	S_0=\int d^4 x\bar{U}_{p'}(x)\,\hat{O}(P,X)\,U_{p}(x)
	\,,
\end{equation}
where $\hat{O}$ is an operator that depends on the type of process, $P$ and $X$ are operators, and $U_{p}(x)$ is the electron wave function in the presence of a laser field (Volkov's solution) with an asymptotic four-momentum $p$, see Eq.~\eqref{eq:vawe_func}.  See also Appendix \ref{sec:app} for definitions and useful formulae.  The matrix element $S$ can be represented as a sum $S=S_1+S_2$, where $S_1$ and $S_2$ correspond to photon emission from the initial and final states, respectively. We neglect the contribution of diagrams related to radiation from intermediate states. The matrix elements $S_{1,2}$ have the following form:
\begin{align}\label{eq:M12}
	S_1&=e\int d^4 x\bar{U}_{p'}(x)\,\hat{O}(P,X)\frac{1}{\hat{\Pi}(\Phi)^2-m^{2}+i0}[\hat{\Pi}(\Phi)+m]e^{ikX}\hat{e}^{*}U_{p}(x)\,,\nonumber\\
	S_2&=e\int d^4 x\bar{U}_{p'}(x)\,e^{ikX}\hat{e}^{*}[\hat{\Pi}(\Phi)+m]\frac{1}{\hat{\Pi}(\Phi)^2-m^{2}+i0}\hat{O}(P,X)U_{p}(x)\,,
\end{align}
where operator $\hat{O}$ is the same as in Eq.~\eqref{eq:M0}, $e$ is the electron charge,  $e^\mu$ and $k^\mu$ are photon polarization and momentum vectors, and $\Pi^{\mu}(\Phi)=P^{\mu}-eA^{\mu}(\Phi)$ with $P^{\mu}=i\partial^{\mu}$.

In our previous work~\cite{Krachkov24}, we obtained expressions enabling us to rewrite the matrix elements $S_{1,2}$ in  terms of $S_0$ as follows:
\begin{equation}\label{eq:prev_res}
	\begin{split}
		 & \frac{1}{\hat{\Pi}(\Phi)-m+i0}\hat{e}^{*}e^{ikX}U_{p}(x)= -i\int_{-\infty}^{\phi}\frac{d\varphi}{p_{-}}e^{-i\int_{\varphi}^{\phi}\frac{d\varphi'}{p_{-}-k_-}(\pi_{p}(\varphi')k)}(\pi_{p}(\varphi) e^*)e^{ikX}U_{p}(x)\,,\\
		&\bar{U}_{p}(x)e^{ikX}\hat{e}^{*}\frac{1}{\hat{\Pi}(\Phi)-m+i0}=-i\int_{\phi}^{+\infty}\frac{d\varphi}{p_{-}}e^{i\int_{\phi}^{\varphi}\frac{d\varphi'}{p_{-}+k_-}(\pi_{p}(\varphi')k)}(\pi_{p}(\varphi) e^*)e^{ikX}\bar{U}_{p}(x)\,,\\
		& \frac{1}{\hat{\Pi}(\Phi)-m+i0}\hat{e}^{*}e^{ikX}V_{p}(x)=-i\int_{\phi}^{+\infty}\frac{d\varphi}{p_{-}}e^{-i\int_{\phi}^{\varphi}\frac{d\varphi'}{p_{-}+k_-}(\pi_{-p}(\varphi')k)}(\pi_{-p}(\varphi) e^*)e^{ikX}V_{p}(x)\,,\\
		& \bar{V}_{p}(x)e^{ikX}\hat{e}^{*}\frac{1}{\hat{\Pi}(\Phi)-m+i0}=-i\int_{-\infty}^{\phi}\frac{d\varphi}{p_{-}}e^{i\int_{\varphi}^{\phi}\frac{d\varphi'}{p_{-}-k_-}(\pi_{-p}(\varphi')k)}(\pi_{-p}(\varphi) e^*)e^{ikX}\bar{V}_{p}(x)\,.
	\end{split}
\end{equation}

These expressions significantly simplify the calculations, but they have some disadvantages. If we try to consider the emission of two soft photons, we cannot use these factors directly, because the phase $\phi$ is included not only in the wave function $U(x)$ but also in the phase factor. Additionally, the extra factor $e^{ikX}$  must be taken into account.  These disadvantages greatly complicate the use of this approximation.
It is easy to show that we can rewrite these expressions in the following manner:
\begin{align}\label{eq:all_res}
	 & \frac{1}{\hat{\Pi}(\Phi)-m+i0}\hat{e}^{*}e^{ikX}U_{p}(x)= -i\int_{-\infty}^{\phi}\frac{d\varphi}{p_{-}}e^{i\int_{0}^{\varphi}\frac{d\varphi'}{{\cal P}_-(p,-k)}(\pi_{p}(\varphi')k)}(\pi_{p}(\varphi)e^{*})U_{{\cal P}(p,-k)}(x)=F_{p}^{(-)}(\phi)U_{{\cal P}(p,-k)}(x)\,,\nonumber\\
		& \bar{U}_{p}(x)e^{ikX}\hat{e}^{*}\frac{1}{\hat{\Pi}(\Phi)-m+i0}=-i\int_{\phi}^{+\infty}\frac{d\varphi}{p_{-}}e^{i\int_{0}^{\varphi}\frac{d\varphi'}{{\cal P}_-(p,k)}(\pi_{p}(\varphi')k)}(\pi_{p}(\varphi)e^{*})\bar{U}_{{\cal P}(p,k)}(x)=F_{p}^{(+)}(\phi)\bar{U}_{{\cal P}(p,k)}(x)\,,\\
		& \frac{1}{\hat{\Pi}(\Phi)-m+i0}\hat{e}^{*}e^{ikX}V_{p}(x)=-i\int_{\phi}^{+\infty}\frac{d\varphi}{p_{-}}e^{i\int_{0}^{\varphi}\frac{d\varphi'}{{\cal P}(-p,-k)}(\pi_{-p}(\varphi')k)}(\pi_{-p}(\varphi) e^*)V_{{\cal P}(-p,-k)}(x)=G_{p}^{(+)}(\phi)V_{{\cal P}(-p,-k)}(x)\,,\nonumber\\
		& \bar{V}_{p}(x)e^{ikX}\hat{e}^{*}\frac{1}{\hat{\Pi}(\Phi)-m+i0}=-i\int_{-\infty}^{\phi}\frac{d\varphi}{p_{-}}e^{i\int_{0}^{\varphi}\frac{d\varphi'}{{\cal P}_-(-p,k)}(\pi_{-p}(\varphi')k)}(\pi_{-p}(\varphi) e^*)\bar{V}_{{\cal P}(-p,k)}(x)=G_{p}^{(-)}(\phi)\bar{V}_{{\cal P}(-p,k)}(x)\,,\nonumber
	\end{align}
where
\begin{equation}
	{\cal P}^\mu(p,\pm k)=(p \pm k)^\mu\mp n^\mu\frac{(pk)}{(p\pm k)_-}\,.
\end{equation}

Note that the function ${\cal P}$ satisfies the following identities:
\begin{equation}
	{\cal P}({\cal P}(p,\pm k_1),\pm k_2)={\cal P}({\cal P}(p,\pm k_2),\pm k_1)={\cal P}(p,\pm k_1\pm k_2)\,,
\end{equation}
since the  momentum of an electron emitting two photons is independent of the emission order.

The new expressions do not have the disadvantages discussed above. Moreover, they have a clear physical interpretation.  The momentum ${\cal P}^\mu(p,\pm k)$ has the following properties: 
\begin{equation}\label{eq:BP}
	{\cal P}_-(p,\pm k)=(p \pm k)_{-}\,,\quad {\cal P}_{\perp}(p,\pm k)=(p \pm k)_{\perp}\,,\quad {\cal P}_+(p,\pm k)=(p \pm k)_+ \mp \frac{(pk)}{(p \pm k)_-}\,,\quad {\cal P}_\mu(p,\pm k) {\cal P}^\mu(p, \pm k)=m^2\,.
\end{equation}
The first expression of~\eqref{eq:all_res} has a clear physical interpretation. In the real photon emission process in the presence of a plane wave background, with an initial electron momentum $p$ and a photon momentum $k$, the final electron momentum is ${\cal P}^\mu(p,- k)$. Although here the photon emission is only a subprocess of a larger process, the appearance of ${\cal P}^\mu(p,- k)$ is quite natural. Since this subprocess is embedded in a more complex reaction, it is accompanied by an additional phase factor. Note that the other expressions in~\eqref{eq:all_res} have a similar interpretation. They differ from the first one by replacing the electron with a positron, and by the fact that photon emission occurs not only in the initial state but also in the final state.

The soft photon approximation allows us to rewrite the amplitude $S$ in terms of $S_0$ as follows:
\begin{align}\label{eq:resS}
	S_1&=\int d^4 x\bar{U}_{p'}(x)\hat{O}(P,X)U_{{\cal P}(p,-k)}(x)F_{p}^{(-)}(\phi)\,,\nonumber\\
	S_2&=\int d^4 xF_{p'}^{(+)}(\phi)\, \bar{U}_{{\cal P}(p',k)}(x)\hat{O}(P,X)U_{p}(x)\,,\nonumber\\
	S&=S_1+S_2=\int_{-\infty}^{+\infty} d\phi \left(\tilde S_{p',{\cal P}(p,-k)}(\phi) F_{p}^{(-)}(\phi)+ \tilde S_{{\cal P}(p',k),p}(\phi) F_{p'}^{(+)}(\phi)\right)\,,
\end{align}
where  $\tilde S_{p',p}(\phi)$ is defined as:
\begin{equation}\label{eq:M00}
	S_0=\int d^4 x\bar{U}_{p'}(x)\,\hat{O}(P,X)\,U_{p}(x)=\int_{-\infty}^{+\infty} d\phi \tilde S_{p',p}(\phi) \,.
\end{equation}

The main feature of the soft photon approximation is that it retains the exact phase in the matrix element, while only the prefactor is approximated. Precise treatment of $k$ in the phase is very important even in the case of $\omega\ll \varepsilon$ due to the integration over the phase along the laser pulse length. The difference in phase on the order of unity significantly changes the probability. The exact accounting for the phase factor enables us to obtain the result with an accuracy of $\mathcal{O}(\frac{\omega}{\varepsilon_{char}})$.

It is worth noting that this result~\eqref{eq:resS} satisfies QED gauge invariance. If we replace the photon polarization vector $e^*$ with the photon momentum $k$, the integrand of $F^{(\pm)}$ becomes a total derivative. This can be easily seen using Eq.~\eqref{eq:prev_res}. Thus, the part from $F^{(+)}$ has the opposite sign in comparison with $F^{(-)}$, and they cancel each other.

It is interesting to compare this result with that of classical radiation. The theory of classical radiation  by a free electron in a plane wave laser field is well established \cite{landau1975classical,Ritus1985,Baier1998}. However, in our case, we consider the emission of a soft photon in the presence of a hard subprocess. This requires a modification of the standard approach, which we develop below. Suppose we have an amplitude $S_0$~\eqref{eq:M00} for a hard subprocess with initial and final  electron momenta  $p$ and $p'$. This amplitude has the following meaning. The quantity $\tilde S_{p',p}(\phi)$ is an amplitude for the process occurring at a specific phase $\phi$. The total amplitude $S_0$ is the integral of this quantity over all possible $\phi$. Let us assume that the hard process occurs instantaneously at some phase $\phi=\phi_0$.

In the classical limit, we assume that the formation length of the photon emission is much larger than the formation length of  the hard subprocess. In such a case, we can assume that the emission of a photon occurs before or after the hard subprocess, i.e., before or after $\phi_0$. When describing the radiation, we should assume that the electron change momentum instantaneously at phase $\phi_0$. The radiation that arises from an instantaneous change in the momentum of a charged particle is precisely classical radiation. The electron momentum as function of phase $\phi$ has the following form:
\begin{equation}
p^\mu(\phi)=\pi_p^\mu(\phi) H(\phi_0-\phi)+\pi_{p'}^\mu(\phi) H(\phi-\phi_0)\,,
\end{equation}
where $\pi_p^\mu(\phi)$ is the classical electron momentum in the presence of a laser field with asymptotic momentum $p$, and $H(x)$ is a Heaviside step function.  The Fourier component of the classical current for the charged particle with momentum $p^\mu(\phi)$ is~\cite{landau1975classical}:
\begin{equation}
j_{\text{cl}}^\mu(k)=-i\int_{-\infty}^{\phi_0}\frac{d\varphi}{p_{-}}e^{i\int_{0}^{\varphi}\frac{d\varphi'}{p_{-}}(\pi_{p}(\varphi')k)}\pi_{p}^\mu(\varphi)-i\int_{\phi_0}^{+\infty}\frac{d\varphi}{p'_{-}}e^{i\int_{0}^{\varphi}\frac{d\varphi'}{p'_{-}}(\pi_{p'}(\varphi')k)}\pi_{p'}^\mu(\varphi)\,.
\end{equation}
In the classical current approximation, the matrix element of the process $i\to f\gamma$ is:
\begin{equation}\label{eq:classical}
S_{\text{cl}}=\int_{-\infty}^{+\infty} d\phi \tilde S_{p',p}(\phi) (j_{\text{cl}}(k)e^*)\,.
\end{equation}
This classical current approximation is mathematically equivalent to the eikonal approximation for soft emission derived using a diagrammatic approach in~\cite{ilderton2013scattering}. 

It is well known that classical radiation does not account for the quantum recoil effect. For this reason,  in~\eqref{eq:classical} $\tilde S_{p',p}(\phi)$ depends on the initial and final electron momenta $p$ and $p'$, without the intermediate momentum shift seen in the soft photon approximation. The phase of the classical current $j_{\text{cl}}^\mu(k)$ is linear in $k$, while the soft photon result has $k_-$ in the phase denominator. These two differences between the soft photon approximation and the classical current approximation lead to a significant discrepancies in numerical results. We will demonstrate this in the next section for several examples.

It is interesting to compare the obtained results with the classical photon emission \cite{LL4}. If we neglect $k_-$ in comparison with $p_-$ in the phase, the result in \eqref{eq:all_res} will coincide with the classical one \cite{LL4}, obtained by discontinuously changing current density four-vector. Note that in the approximation under consideration, we neglect terms proportional to $k_\mu$ only in the  preexponential factors. We calculate the phase multiplier exactly.This distinction is physically vital: because the phase is integrated over the macroscopic laser pulse length, even a small $k$
can produce a phase difference on the order of unity, which fundamentally reshapes the interference pattern and the final probability.

The comparison between the classical and the soft photon approximations leads to the understanding of new important features of soft photon approximation. The soft photon factors $F_p^{\pm}$ and $G_p^{\pm}$ do not depend on the spin of the charged particle. As is well known, the phase of the wave function in a plane wave background does not depend on the spin of the particle, as it is a quasiclassical phase given by the classical action, which is spin-independent. Furthermore, the preexponential factors in the soft photon approximation coincide with the prefactor of the classical result, which is also spin-independent. Thus, the soft photon approximation factors $F_p^{\pm}$ and $G_p^{\pm}$ are spin independent.

For completeness, we consider the emission of two soft photons with momenta $k_{1,2}$  and with the polarization vectors $e_{1,2}$ from an initial electron

\begin{align}\label{eq:2photon}
		& \frac{1}{\hat{\Pi}(\Phi)-m+i0}\hat{e_2}^{*}e^{ik_2X} \frac{1}{\hat{\Pi}(\Phi)-m+i0}\hat{e_1}^{*}e^{ik_1X}U_{p}(x)=\nonumber\\
		 &\frac{1}{\hat{\Pi}(\Phi)-m+i0}\hat{e_2}^{*}e^{ik_2X}(-i)\int_{-\infty}^{\phi}\frac{d\varphi_1}{p_{-}}e^{i\int_{0}^{\varphi_1}\frac{d\varphi'_1}{{\cal P}_-(p,-k_1)}(\pi_{p}(\varphi'_1)k_1)}(\pi_{p}(\varphi_1)e_1^{*})U_{{\cal P}(p,-k_1)}(x)=\\
 		 &(-i)\int_{-\infty}^{\phi}\frac{d\varphi_2}{p_{-}}e^{i\int_{0}^{\varphi_2}\frac{d\varphi'_2}{{\cal P}_-(p,-k_1-k_2)}(\pi_{{\cal P}(p,-k_1)}(\varphi'_2)k_2)}(\pi_{p}(\varphi_2)e_2^{*})\nonumber\\
 		 &\times(-i)\int_{-\infty}^{\varphi_2}\frac{d\varphi_1}{p_{-}}e^{i\int_{0}^{\varphi_1}\frac{d\varphi'_1}{{\cal P}_-(p,-k_1)}(\pi_{p}(\varphi'_1)k_1)}(\pi_{p}(\varphi_1)e_1^{*})
 		 U_{{\cal P}(p,-k_1-k_2)}(x)\,.\nonumber
\end{align}

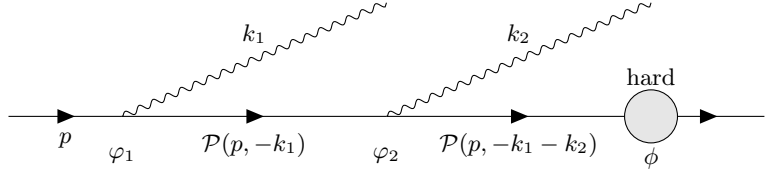
\begin{figure}
	\centering
	\begin{tikzpicture}
		\begin{feynman}
			\vertex (i);
			\vertex [right=1.5cm of i] (v1);
			\vertex [right=3.5cm of v1] (v2);
			\vertex [right=3.5cm of v2] (blob);
			\vertex [right=1.5cm of blob] (o);
			\vertex [above=1.5cm of v2] (p1);
			\vertex [above=1.5cm of blob] (p2);
			\draw [fermion] (i) -- node[below=0.1cm] {\(p\)} (v1);
			\draw [fermion] (v1) -- node[below=0.1cm] {\({\cal P}(p,-k_1)\)} (v2);
			\draw [fermion] (v2) -- node[below=0.1cm] {\({\cal P}(p,-k_1-k_2)\)} (blob);
			\draw [fermion] (blob) -- (o);
			\draw [photon] (v1) -- node[above=0.1cm] {\(k_1\)} (p1);
			\draw [photon] (v2) -- node[above=0.1cm] {\(k_2\)} (p2);
			\node[draw, circle, minimum size=0.7cm, fill=gray!20, inner sep=0pt] at (blob) {};
			\node[below=0.3cm of v1] {\(\varphi_1\)};
			\node[below=0.3cm of v2] {\(\varphi_2\)};
			\node[below=0.3cm of blob] {\(\phi\)};
			\node[above=0.3cm of blob] {\(\text{hard}\)};
		\end{feynman}
	\end{tikzpicture}
	\caption{Space-time picture of double soft photon emission from an initial electron line.}
	\label{fig:2photon}
\end{figure}

For a detailed understanding of this identity, see the derivation of the soft photon approximation~\cite{Krachkov24}. The result has a clear physical picture, depicted in Fig.~\ref{fig:2photon}. The initial electron with momentum $p$ first emits a photon with momentum $k_1$ at the point $\varphi_1$. Thereafter, this electron propagates with the momentum ${\cal P}(p,-k_1)$. It then emits the second photon at the point $\varphi_2$, while changing the momentum to ${\cal P}(p,-k_1-k_2)$. These two integrals in~\eqref{eq:2photon} in some sense correspond to propagators, i.e., in the absence of a plane-wave laser field, these integrals exactly correspond to the denominator like $(p k_1)$. The numerator of the phase contains the momentum before the soft photon emission, while the denominator of the phase contains the electron momentum after the soft photon emission. Also note that the action of the second operator on the first soft photon factor results in a shift of the upper limit of integration. It is clear from this picture that phases satisfy the following inequality $\varphi_1<\varphi_2<\phi$, which corresponds to the integration region in~\eqref{eq:2photon}. Keeping this simple physical picture in mind,  it is easy to generalize this to $n$ soft photon emission amplitudes, as well as to other  in/out or electron/positron configurations.

\section{$N \gamma$ nonlinear Compton scattering}
\subsection{Single Compton scattering}\label{sec:SCS}
We start with the simplest example: nonlinear Compton scattering. Although this process was discussed in the framework of the soft-photon approximation in~\cite{Krachkov24}, we revisit it here for the following reasons. Nonlinear Compton scattering is the simplest example of a  QED process in a plane-wave background field.  It has been extensively studied; see, e.g., Refs.~\cite{Harvey2009,Narozhnyi1996,Ritus1985,PhysRev.133.A705,PhysRevD.99.096018,Seipt2013,Ivanov2004,Zeldovich,NikiRitus,Mackenroth2011,Goldman1969}. By comparing the exact and the soft photon results, we can understand the region of applicability and the main features of the soft photon approximation.

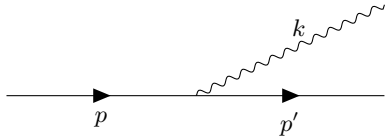
\begin{figure}[ht]
	\centering
	\begin{tikzpicture}
		\begin{feynman}
			\vertex (i);
			\vertex [right=2.5cm of i] (v);
			\vertex [right=2.5cm of v] (o);
			\vertex [above=1.2cm of o] (p1);
			\draw [fermion] (i) -- node[below=0.1cm] {\(p\)} (v);
			\draw [fermion] (v) -- node[below=0.1cm] {\(p'\)} (o);
			\draw [photon] (v) -- node[left=0.3cm, pos=0.75] {\(k\)} (p1);
		\end{feynman}
	\end{tikzpicture}
	\caption{Feynman diagram for single nonlinear Compton scattering.}
	\label{pic:Compton_diagram}
\end{figure}

The Feynman diagram for nonlinear Compton scattering is shown in Fig.~\ref{pic:Compton_diagram}, and the corresponding matrix element is:
\begin{align}
	S&=\int d^4x\bar{U}_{p'}(x)e^{ikX}\hat{e}^{*}U_{p}(x)\,,\nonumber
\end{align}
where $p$, $p'$, and $k$ are momenta of the initial electron, final electron, and emitted photon, respectively. $e^\mu$ is the photon polarization vector. As mentioned above, $p$, $p'$, and $k$ satisfy the following relations:
\begin{equation}
	{\cal P}(p,- k)=p'\,,\quad{\cal P}(p', k)=p\,.
\end{equation}

At first glance, it seems that the soft photon approximation is not applicable to this process. Nevertheless, we can apply this approximation using the following trick. We can write the identity operator in the form:
\begin{equation}
	\mathbb{1}=[\hat{\Pi}(\Phi)-m] \frac{1}{\hat{\Pi}(\Phi)^2-m^{2}+i0}[\hat{\Pi}(\Phi)+m]\,.
\end{equation}

Note that all operators should act on the right side; otherwise, we would obtain an incorrect result. This is crucial because the operator $[\hat{\Pi}(\Phi)-m]$, when acting on the Volkov state $\bar{U}_{p'}(x)$ from the left, gives zero, which is a direct consequence of the Dirac equation.

\begin{align}\label{eq:res_S}
	S=&\int d^4x\bar{U}_{p'}(x)[\hat{\Pi}(\Phi)-m] \frac{1}{\hat{\Pi}(\Phi)^2-m^{2}+i0}[\hat{\Pi}(\Phi)+m]  e^{ikX}\hat{e}^{*}U_{p}(x) \nonumber\\ 
	=&\int d^4x\bar{U}_{p'}(x)[\hat{\Pi}(\Phi)-m] 	(-i)\int_{-\infty}^{\phi}\frac{d\varphi}{p_{-}}e^{i\int_{0}^{\varphi}\frac{d\varphi'}{p_{-}-k_{-}}(\pi_{p}(\varphi')k)}(\pi_{p}(\varphi)e^{*})U_{p'}(x) \nonumber\\ 
	=&\int d^4x\bar{U}_{p'}(x)P_\phi\hat{n}	(i)\int_{-\infty}^{\phi}\frac{d\varphi}{p_{-}}e^{i\int_{0}^{\varphi}\frac{d\varphi'}{p_{-}-k_{-}}(\pi_{p}(\varphi')k)}(\pi_{p}(\varphi)e^{*})U_{p'}(x) \nonumber\\ 
	=&\int d^4x	\frac{\bar{U}_{p'}(x)\hat{n}U_{p'}(x)}{p_{-}}e^{i\int_{0}^{\phi}\frac{d\phi'}{p_{-}-k_{-}}(\pi_{p}(\phi')k)}(\pi_{p}(\phi)e^{*}) \nonumber\\ 
	=&2(2\pi)^3 \delta(\bm p_\perp-\bm p'_\perp-\bm k_\perp) \delta(p_-- p'_-- k_-)\int_{-\infty}^{+\infty}d\phi (\pi_p(\phi) e^*) e^{i\int_{0}^{\phi}\frac{d\varphi'}{p'_{-}}(\pi_{p}(\varphi')k)}\,.
\end{align}
Note that for consistency and completeness of the theory, this expression should be obtained when the operators act on both sides. In order to obtain the  result when the operator acts on the left side, we should write the identity operator as:
\begin{equation}
	\mathbb{1}= [\hat{\Pi}(\Phi)+m]\frac{1}{\hat{\Pi}(\Phi)^2-m^{2}+i0}[\hat{\Pi}(\Phi)-m]\,.
\end{equation}
The final result will be the same, as expected, if we take into account the following phase identity: 
\begin{equation}\label{eq:phase_id}
	\frac{\pi_{p}(\varphi')k}{{\cal P}_-(p,- k)}=\frac{\pi_{{\cal P}(p,- k)}(\varphi')k}{p_-}\,.
\end{equation}

The electron-emission  energy spectrum  $\dfrac{d E}{d^3k}$ is given by:
\begin{align}\label{eq:res_CS}
	\dfrac{d E}{d^3 k}&=\frac{\alpha m^2}{4\pi^2 p_-^2}\left[\xi^2(|f_1|^2-\operatorname{Re} \, f_0 f_2^*)-|f_0|^2\right]\,,\\
	f_i&=\int_{-\infty}^{+\infty} d\phi\left(\frac{e \bm A(\phi)}{m\xi}\right)^i e^{i\int_{0}^{\phi}\frac{d\phi'}{p'_{-}}(\pi_{p}(\phi')k)}\,,\nonumber
\end{align}
where $\xi=|e| E_0/m\omega_0$, with $E_0$ and $\omega_0$ being the laser electric field amplitude and angular frequency.
The energy spectrum for nonlinear Compton scattering \eqref{eq:res_CS} coincides with the classical result \cite{Ritus1985}  if we replace $p'_-\rightarrow p_-$ in the phase of $f_i$. On the other hand, the result \eqref{eq:res_CS} coincides with the exact energy spectrum for nonlinear Compton scattering \cite{Mackenroth2011}, to leading order in the parameter $\omega/\varepsilon$ in the preexponential factor and exactly in the phase factor.

To evaluate the accuracy of this approximation, we now compare the soft photon result with both the classical and exact quantum calculations for specific kinematic configurations. The problem has a large parameter space – one could vary the laser pulse shape, duration, intensity, and collision angle or electron energy. However, we focus on a single configuration that is most relevant experimentally: an ultra-relativistic electron moving along the $z$-direction colliding with a counter-propagating laser pulse. This geometry yields the maximal effective intensity parameter $\xi$ and is the standard setup in modern strong‑field QED experiments \cite{Abramowicz2021,gales2018extreme,yakimenko2019facet}. For such kinematics, the characteristic angle between the emitted photon and the initial electron is small, $\theta\sim 1/\gamma$. Thus, we can use a small-angle expansion and a two-dimensional vector $\bm\theta_{pk}$, where $\bm\theta_{q,l}=\frac{\bm q_\perp}{q}-\frac{\bm l _\perp}{l}$ for arbitrary vectors $\bm q\,,\bm l$. Under these conditions, the spectrum is:
\begin{align}\label{eq:res_HE}
	\dfrac{d E}{d^3 k}&=\frac{\alpha m^2}{(4\pi)^2 \varepsilon^2}|\xi \bm f_1-\frac{\varepsilon \bm \theta_{pk}}{m}   f_0|^2\,,\\
	f_i&=\int_{-\infty}^{+\infty} d\phi\left(\frac{e \bm A(\phi)}{m\xi}\right)^i e^{i\int_{0}^{\phi}\frac{\omega d\varphi }{4\varepsilon(\varepsilon-\omega)}\left[m^{2}+(\varepsilon\bm{\theta}_{pk}-e\bm{A}(\varphi))^{2}\right]}\,,\quad \bm f_1=f_1 \bm e_x\,.\nonumber
\end{align}
Here we use the following identities:
\begin{align}
&k\pi_{p}=\frac{\omega}{2\varepsilon}(m^{2}+(\varepsilon\bm{\theta}_{pk}-e\bm{A})^{2})\,,\quad\pi_{p}e=-(\varepsilon\bm{\theta}_{pk}-e\bm{A})\cdot\bm{e}\,,\nonumber\\
&\int_{-\infty}^{+\infty} d e^{i\int_{0}^{\phi}\frac{\omega d\varphi }{4\varepsilon(\varepsilon-\omega)}\left[m^{2}+(\varepsilon\bm{\theta}_{pk}-e\bm{A}(\varphi))^{2}\right]}=f_0\left(1+\frac{\varepsilon^2\theta_{pk}^2}{m^2}\right)-2\xi\frac{\varepsilon\bm\theta_{pk}}{m}\cdot\bm f_1+\xi^2 f_2=0\,.\nonumber
\end{align}
The spectrum of the classical radiation can be simply obtained from  Eq.~\eqref{eq:res_HE} by neglecting $\omega$ in comparison with $\varepsilon$ in the denominator of the phase of $f_i$.

For completeness, we also present here the exact spectrum $dE_t/d^3k$ of nonlinear Compton scattering, which can be obtained from the general result of Ref.~\cite{Mackenroth2011} in the high-energy limit:
\begin{equation}\label{eq:res_HE_exact}
	\dfrac{d E_t}{d^3 k}=\frac{\alpha m^2}{2(4\pi)^2 \varepsilon^2(\varepsilon-\omega)^2}\left[\omega^2|f_0|^2+\left(\varepsilon^2+(\varepsilon-\omega)^2\right)\left|\xi \bm f_1-\frac{\varepsilon \bm \theta}{m}   f_0\right|^2\right]\,.
\end{equation}

\begin{figure}
	\centering
	\includegraphics[width=0.9\linewidth]{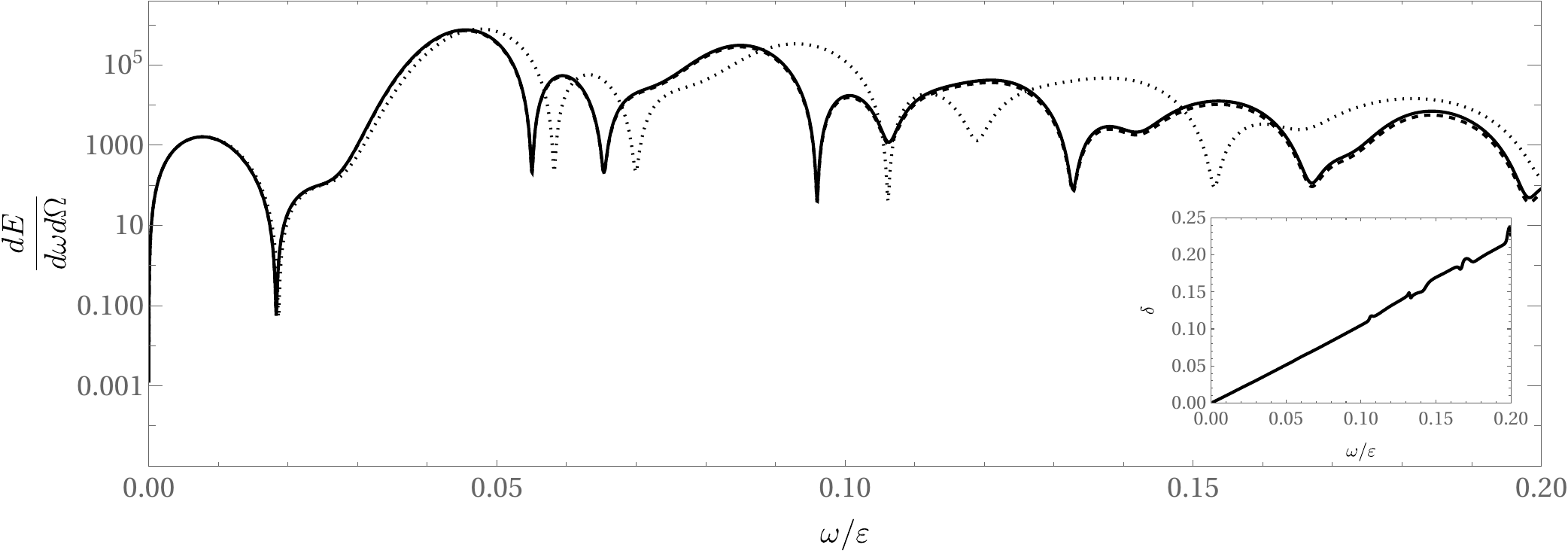}
	\caption{The energy spectrum for nonlinear Compton scattering $\dfrac{d E}{d\omega d\Omega}$ as a function of  $\omega/\varepsilon$, the photon angles are $\theta_{pk,x}=\frac{m}{2\varepsilon}$, $\theta_{pk,y}=\frac{m}{\varepsilon}$. The details of numerical computation are discussed in the text. The black line corresponds to the exact result, the dashed line corresponds to the soft photon approximation result, and the dotted line corresponds to the classical result. Inset: the relative difference $\delta$ between the exact and the soft photon energy spectra for nonlinear Compton scattering as a function of  $\omega/\varepsilon$.}
	\label{pic:Compton}
\end{figure}

In order to show the importance of the exact phase result, we plot the energy spectrum in Fig.~\ref{pic:Compton}. For numerical evaluations, we consider electrons with initial energy $\varepsilon=10^4 m$ in head-on collisions with a linearly polarized laser pulse. Calculations have been performed for $\xi=1$ and a pulse shape  $\bm A(\phi)= \frac{m \xi}{|e|} \cos[\omega_0\phi] g(\phi)\bm e_x$, with the envelope function $g(\phi)=\cos^2(\pi \omega_0 \phi/2\tau)$ for $-\tau\leq\omega_0 \phi\leq \tau$ and zero otherwise, with dimensionless pulse length $\tau=20$ and $\omega_0=1.55$ eV. We also assume that the electron charge is negative, i.e., $e=-|e|$. We set $m=1$, thus all numerical quantities are presented in these units.

We see from Fig.~\ref{pic:Compton} that the classical result coincides with the exact one only in a narrow region $\omega/\varepsilon\lesssim0.01$, while the soft photon approximation well fits the exact result in a wider range. The reason for the significant discrepancy between the classical and the exact quantum results can be simply explained as follows. Although the integrand in the phase differs slightly in the classical and quantum results, the integral over a wide range for a long laser pulse leads to a difference in the phases on the order of unity. The difference in the phases on the order of unity leads to a significant difference in the spectrum. The phase in the soft photon approximation is exactly the same as in the exact result, resulting in significantly better accuracy compared to the classical result.

Note that Fig.~\ref{pic:Compton} has a logarithmic scale, so the difference between the exact and the soft photon results seems small. As can be seen from Fig.~\ref{pic:Compton},  the relative difference between the exact and the soft photon results reaches $25\%$ at $\frac{\omega}{\varepsilon}=0.2\,$.  The presence of peaks in Fig.~\ref{pic:Compton} is due to the fact that the functions $f_i$ have different resonant frequencies $\omega$.

\begin{figure}
	\centering
	\includegraphics[width=0.9\linewidth]{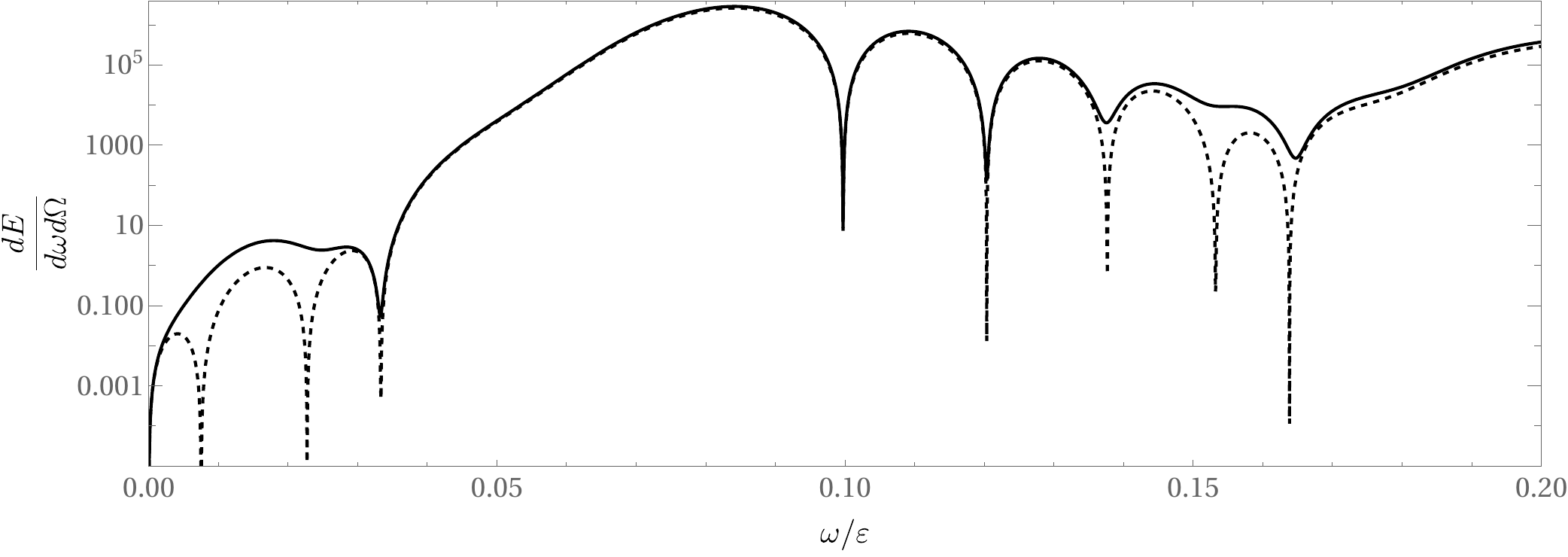}
	\caption{The energy spectrum for nonlinear Compton scattering $\frac{d E}{d\omega d\Omega}$ as a function of  $\omega/\varepsilon$, the photon angles are $\theta_{pk,x}=\theta_{pk,y}=0$. The details of numerical computation are discussed in the text. The black line corresponds to the exact result, the dashed line corresponds to the soft photon approximation result.}
	\label{pic:Compton00}
\end{figure}

\begin{figure}
	\centering
	\includegraphics[width=0.7\linewidth]{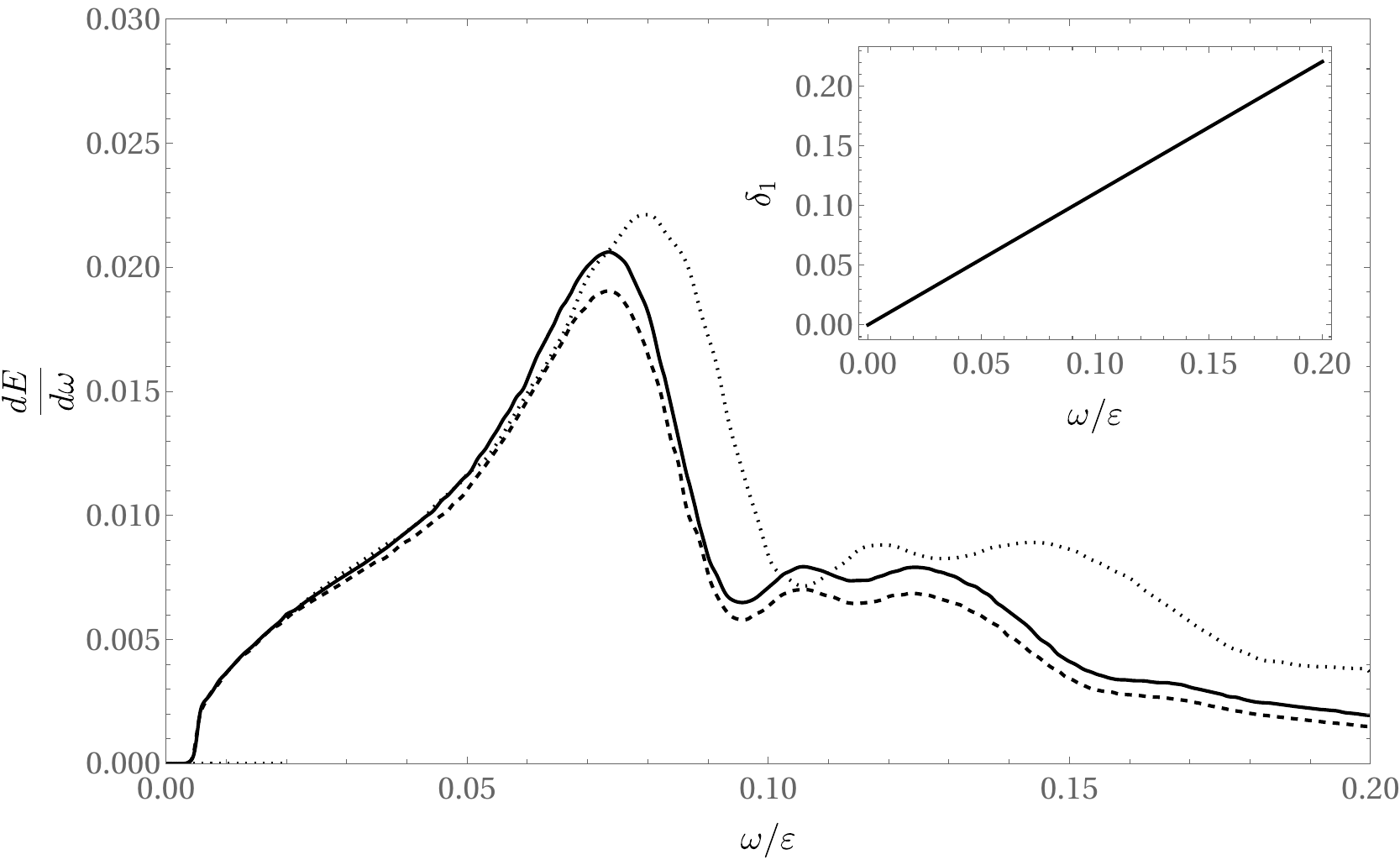}
	\caption{The energy spectrum for nonlinear Compton scattering $\frac{d E}{d\omega}$ as a function of  $\omega/\varepsilon$. The details of numerical computation are discussed in the text. The black line corresponds to the exact result, the dashed line corresponds to the soft photon approximation result, and the dotted line corresponds to the classical result. Inset: the relative difference $\delta_1$ between the exact and the soft photon energy spectra as a function of  $\omega/\varepsilon$.}
	\label{pic:IntCompton}
\end{figure}

We see from Fig.~\ref{pic:Compton00} that the difference between the exact and the soft photon results is huge at $\theta_{pk,x}=\theta_{pk,y}=0$. This point is special, because at this point the soft photon spectrum~\eqref{eq:res_HE} depends on $f_1$ and does not depend on $f_0$, while the exact spectrum~\eqref{eq:res_HE_exact} depends on both $f_0$ and $f_1$. In the parametric region where $\omega f_0\gg f_1$, there is a significant difference between the exact and the soft photon results. Note that such a huge difference between these two spectra occurs only in a very narrow region of the  parameter $\bm\theta_{pk}$, i.e., $\frac{\varepsilon}{m}\theta_{pk}\sim\frac{\omega}{\varepsilon}\ll1$. Thus, in the integrated spectrum $\dfrac{d E}{d\omega}$, the difference between the exact and soft photon results is $\mathcal{O}\left(\frac{\omega}{\varepsilon}\right)$. To illustrate this statement, we show the spectrum integrated over the photon angles $\frac{d E}{d\omega}$ as a function of  $\omega/\varepsilon$. We see from Fig.~\ref{pic:IntCompton} that the integrated spectrum $\frac{d E}{d\omega}$ has the same accuracy as $\frac{d E}{d\omega d\Omega}$ in the soft photon approximation.
Note that the classical result significantly differs from the exact result at $\omega/\varepsilon>0.07$.

\subsection{Double Compton scattering}\label{sec:DCS}
The nonlinear double Compton scattering is a more complicated process in comparison with single Compton scattering. There is an electron in an intermediate state in this process, which results in new phenomena. The electron in the intermediate state can be on the mass shell; this leads to a cascade, where the process divides into two independent parts. In such a case, the probability of double Compton scattering increases quadratically with respect to the length of the laser pulse for the long pulse case. From the mathematical point of view, double Compton scattering is also more complicated than single Compton scattering, because the matrix element of double Compton scattering includes an electron propagator, which is more complicated than the wave function. Moreover, the matrix element of double Compton scattering has a two-dimensional integral over highly oscillating functions, which is hard to compute. In spite of the fact that this process is widely discussed in~\cite{mackenroth2013nonlinear,PhysRevD.99.096018, seipt2012two,de2024production}, a large number of parameters in this process require further investigation. The soft photon approximation can help with this problem because it significantly simplifies the matrix element calculation.

In the previous section, we have performed a detailed comparison between the soft photon approximation, the classical current approximation, and the exact quantum result for single nonlinear Compton scattering. This comparison has established the domain of applicability of the method: the soft photon approximation reproduces the exact spectrum  with an accuracy of $\mathcal{O}(\omega/\varepsilon)$, while the classical result is valid only in a narrow region. 
The improved accuracy stems from the exact treatment of the phase, which correctly captures the formation length physics over the extended laser pulse. Having verified the method on this benchmark process, we now turn to double Compton scattering, which is significantly more computationally demanding. In the present work, we do not perform the exact calculation of nonlinear double Compton scattering, due to its complexity. Instead, the goal of this paper  is to demonstrate the power of the soft photon approximation as a practical tool: we show that it allows one to obtain the complete analytical result for a process of this complexity in a straightforward and computationally efficient manner. Accordingly, in what follows we present double Compton scattering probability within the soft photon approximation and analyze its main physical features.

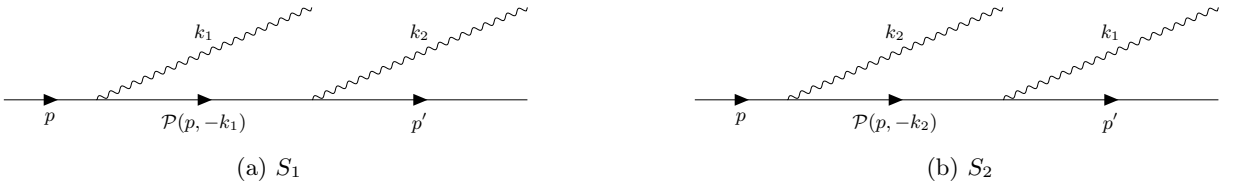
\begin{figure}[htbp]
	\centering
	\begin{minipage}{0.49\textwidth}
		\centering
		\resizebox{0.8\textwidth}{!}{%
		\begin{tikzpicture}
		\begin{feynman}
	\vertex (i);
	\vertex [right=1.5cm of i] (v1);
	\vertex [right=3.5cm of v1] (v2);
	\vertex [right=3.5cm of v2] (blob);
	\vertex [above=1.5cm of v2] (p1);	
	\draw [fermion] (i) -- node[below=0.1cm] {\(p\)} (v1);
	\draw [fermion] (v1) -- node[below=0.1cm] {\({\cal P}(p,-k_1)\)} (v2);
	\draw [fermion] (v2) -- node[below=0.1cm] {\(p'\)} (blob);
	\draw [photon] (v1) -- node[above=0.1cm] {\(k_1\)} (p1);
	\draw [photon] (v2) -- node[above=0.1cm] {\(k_2\)} (p2);	
		\end{feynman}
		\end{tikzpicture}
	}
		\subcaption{$S_1$}
		\label{fig:S1}
	\end{minipage}
	\hfill
	\begin{minipage}{0.49\textwidth}
		\centering
				\resizebox{0.8\textwidth}{!}{%
		\begin{tikzpicture}
		\begin{feynman}
	\vertex (i);
	\vertex [right=1.5cm of i] (v1);
	\vertex [right=3.5cm of v1] (v2);
	\vertex [right=3.5cm of v2] (blob);
	\vertex [above=1.5cm of v2] (p1);	
	\draw [fermion] (i) -- node[below=0.1cm] {\(p\)} (v1);
	\draw [fermion] (v1) -- node[below=0.1cm] {\({\cal P}(p,-k_2)\)} (v2);
	\draw [fermion] (v2) -- node[below=0.1cm] {\(p'\)} (blob);
	\draw [photon] (v1) -- node[above=0.1cm] {\(k_2\)} (p1);
	\draw [photon] (v2) -- node[above=0.1cm] {\(k_1\)} (p2);
\end{feynman}
		\end{tikzpicture}
	}
		\subcaption{$S_2$}
		\label{fig:S2}
	\end{minipage}
	\caption{Feynman diagrams for nonlinear double Compton scattering.}
	\label{pic:DC}
\end{figure}

Double Compton scattering has two Feynman diagrams, which are depicted in Fig.~\ref{pic:DC}.

We will discuss the process where the initial and the final electrons have momenta $p$ and $p'$, respectively. The two emitted photons have momenta and polarization vectors $k_{1,2}$ and $e_{1,2}$, as shown in Fig.~\ref{pic:DC}. Note that, according to the definition~\eqref{eq:BP}, momenta $p$, $p'$, and $k_{1,2}$ satisfy the following identities:
\begin{equation}
	p'={\cal P}(p,-k_1-k_2)\,,\quad p={\cal P}(p',k_1+k_2)\,,\quad {\cal P}(p,-k_{1,2})={\cal P}(p',k_{2,1})  \,.
\end{equation}

The matrix element for double Compton scattering is $S=S_1+S_2$, where:
\begin{align}
	S_1=&\int d^4x\bar{U}_{p'}(x)e^{ik_2X}\hat{e_2}^{*} \frac{1}{\hat{\Pi}(\Phi)^2-m^{2}+i0}[\hat{\Pi}(\Phi)+m]  e^{ik_1X}\hat{e_1}^{*}U_{p}(x)\,,\nonumber\\
	S_2=&\int d^4x\bar{U}_{p'}(x)e^{ik_1X}\hat{e_1}^{*} \frac{1}{\hat{\Pi}(\Phi)^2-m^{2}+i0}[\hat{\Pi}(\Phi)+m]  e^{ik_2X}\hat{e_2}^{*}U_{p}(x)\,.
\end{align}

Let the photon with momentum $k_2$ be soft. We can apply Eqs.~\eqref{eq:all_res} to the photon with momentum $k_2$:
\begin{align}\label{eq:DC1}
	S_1=&-i\int d^4x\int_{\phi}^{+\infty}\frac{d\varphi}{p'_{-}}e^{i\int_{0}^{\varphi}\frac{d\varphi'}{{\cal P}_-(p',k_2)}(\pi_{p'}(\varphi')k_2)}(\pi_{p'}(\varphi)e_2^{*})\bar{U}_{{\cal P}(p',k_2)}(x)  e^{ik_1X}\hat{e_1}^{*}U_{p}(x)\,,\nonumber\\
	S_2=&-i\int d^4x\bar{U}_{p'}(x)e^{ik_1X}\hat{e_1}^{*} \int_{-\infty}^{\phi}\frac{d\varphi}{p_{-}}e^{i\int_{0}^{\varphi}\frac{d\varphi'}{{\cal P}_-(p,-k_2)}(\pi_{p}(\varphi')k_2)}(\pi_{p}(\varphi)e_2^{*})U_{{\cal P}(p,-k_2)}(x)\,.
\end{align}
Thus, both matrix elements can be expressed through the matrix element of nonlinear Compton scattering, but with different initial and final momenta. This result can be used in the case where one photon is soft, while the second photon is hard. We will not discuss this kinematics here. 

If we assume that the photon with momentum $k_1$ is also soft, we can apply the soft photon approximation a second time. We can use the result from the previous section, Eq.~\eqref{eq:res_S}. The matrix elements are:
\begin{align}\label{eq:DC2}
	S_1=&-i	2(2\pi)^3 \delta(\bm p_\perp-\bm p'_\perp-\bm k_{1\perp}-\bm k_{2\perp}) \delta(p_-- p'_-- k_{1-}- k_{2-})\int_{-\infty}^{+\infty}d\phi \int_{\phi}^{+\infty}\frac{d\varphi}{p_{-}}\times\nonumber\\
	&e^{i\int_{0}^{\varphi}\frac{d\varphi'}{{\cal P}_-(p,-k_1)}(\pi_{{\cal P}(p,-k_1-k_2)}(\varphi')k_2)}(\pi_{p}(\varphi)e_2^{*}) (\pi_p(\phi) e_1^*) e^{i\int_{0}^{\phi}\frac{d\varphi'}{{\cal P}_{-}(p,-k_{1})}(\pi_{p}(\varphi')k_1)}
	\,,\nonumber\\
	S_2=&-i	2(2\pi)^3 \delta(\bm p_\perp-\bm p'_\perp-\bm k_{1\perp}-\bm k_{2\perp}) \delta(p_-- p'_-- k_{1-}- k_{2-})\int_{-\infty}^{+\infty}d\phi \int_{-\infty}^{\phi}\frac{d\varphi}{p_{-}}\times\nonumber\\
	&e^{i\int_{0}^{\varphi}\frac{d\varphi'}{{\cal P}_-(p,-k_2)}(\pi_{p}(\varphi')k_2)}(\pi_{p}(\varphi)e_2^{*}) (\pi_p(\phi) e_1^*) e^{i\int_{0}^{\phi}\frac{d\varphi'}{{\cal P}_{-}(p,-k_{1}-k_{2})}(\pi_{{\cal P}(p,-k_2)}(\varphi')k_1)}\,.
\end{align}
The matrix element $S_1$ with substitutions $k_1\leftrightarrow k_2$ and $e_1\leftrightarrow e_2$ coincides with $S_2$. This can be seen by using Eq.~\eqref{eq:phase_id}.

Note also that if we neglect $k_{1,2}$ in the phase in comparison with $p$, i.e., replace ${\cal P}(p,-k_{1,2})$ with $p$, we obtain the classical result. In such a case, the amplitude in the classical approximation $S_c=S_{1,c}+S_{2,c}$ is factorized:
\begin{align}\label{eq:DC_clasical}
	S_c=&-i	\frac{2}{p_{-}}(2\pi)^3 \delta(\bm p_\perp-\bm p'_\perp-\bm k_{1\perp}-\bm k_{2\perp}) \delta(p_-- p'_-- k_{1-}- k_{2-}) \times\nonumber\\
	&\left(\int_{-\infty}^{+\infty}d\phi (\pi_p(\phi) e_1^*) e^{i\int_{0}^{\phi}\frac{d\varphi'}{p_-}(\pi_{p}(\varphi')k_1)}\right)
	\left(\int_{-\infty}^{+\infty}d\phi (\pi_p(\phi) e_2^*) e^{i\int_{0}^{\phi}\frac{d\varphi'}{p_-}(\pi_{p}(\varphi')k_2)}\right)
	\,.
\end{align}
The soft photon approximation takes into account the change of the electron momentum after the soft photon emission in the phase, which corresponds to replacing $p$ with ${\cal P}(p,-k_{1,2})$. This replacement breaks the factorization, as the intermediate electron momenta in the two emission amplitude are different. This is a key difference between the soft photon and classical approximations: the former captures the quantum recoil effect, while the latter does not.

The result~\eqref{eq:DC2} for the matrix element of nonlinear double Compton scattering can be used for arbitrary kinematics and field configurations. We focus our attention on one configuration: an ultra-relativistic electron in head-on collision with a laser pulse.  We will use the same kinematics and field configuration as in Sec.~\ref{sec:SCS}. In this case, the matrix elements $S_{1,2}$ can be expressed in terms of integrals: 
\begin{align}\label{eq:fgij}
	f_{i,j}&=\int_{-\infty}^{+\infty}d\phi \int_{-\infty}^{\phi}d\varphi \left(\frac{e \bm A(\phi)}{m\xi}\right)^i\left(\frac{e \bm A(\varphi)}{m\xi}\right)^j  e^{i \Phi_1(\varphi) +i \Phi_{21}(\phi)}\,,\nonumber\\
	g_{i,j}&=\int_{-\infty}^{+\infty}d\phi \int_{-\infty}^{\phi}d\varphi \left(\frac{e \bm A(\phi)}{m\xi}\right)^i\left(\frac{e \bm A(\varphi)}{m\xi}\right)^j  e^{i \Phi_2(\varphi) +i \Phi_{12}(\phi)}\,,\\
	\Phi_i(\varphi)&= \int_0^\varphi d\varphi' \frac{(\pi_{p}(\varphi')k_{i})}{{\cal P}_{-}(p,-k_{i})} 
	=\omega_i \int_0^\varphi d\varphi'\frac{\left(m^{2}+(\varepsilon\bm{\theta}_{pk_i}-e\bm{A})^{2}\right)}{4\varepsilon(\varepsilon-\omega_i)}\,\nonumber\\
	\Phi_{i,j}(\varphi)&= \int_0^\varphi d\varphi'\frac{(\pi_{{\cal P}(p,-k_j)}(\varphi')k_i)}{{\cal P}_-(p,-k_i-k_j)} = \omega_i\int_0^\varphi d\varphi'\frac{\left(m^{2}+((\varepsilon-\omega_j)\bm{\theta}_{pk_i}+\omega_j \bm{\theta}_{pk_j}-e\bm{A})^{2}\right)}{4(\varepsilon-\omega_j)(\varepsilon-\omega_i-\omega_j)}\,.\nonumber
\end{align}
The functions $f_{ij}$ correspond to the matrix element $S_1$, while the functions $g_{ij}$ correspond to the matrix element $S_2$.

The probability of two photon emission is:
\begin{align}\label{eq:W}
	&\dfrac{d W}{d^3 k_1 d^3 k_2}=\frac{\alpha^2 m^4}{(4\pi \varepsilon)^4\omega_1\omega_2} \left(A_{11}+A_{22}+2\Re A_{12} \right) \,, \\
	A_{11}&=\int_{-\infty}^{+\infty}d\phi_1d\phi_2 \int_{-\infty}^{\phi_1}d\varphi_1\int_{-\infty}^{\phi_2}d\varphi_2 e^{i \Phi_1(\varphi_1) +i \Phi_{21}(\phi_1)-i \Phi_1(\varphi_2) -i \Phi_{21}(\phi_2)}\times\nonumber\\
	&\left(\frac{\varepsilon}{m}\bm{\theta}_{pk_1}-\xi\bm{A}(\varphi_1)\right)\cdot\left(\frac{\varepsilon}{m}\bm{\theta}_{pk_1}-\xi\bm{A}(\varphi_2)\right) \left(\frac{\varepsilon}{m}\bm{\theta}_{pk_2}-\xi\bm{A}(\phi_1)\right)\cdot\left(\frac{\varepsilon}{m}\bm{\theta}_{pk_2}-\xi\bm{A}(\phi_2)\right)\,,\nonumber\\
	A_{22}& = \Bigl[ A_{11} \Bigr]_{\bm\theta_{pk_1}\leftrightarrow\bm\theta_{pk_2}\,,\;\Phi_1\rightarrow\Phi_2\,,\;\Phi_{21}\rightarrow\Phi_{12}}\nonumber\,,\\
	A_{12}&=\int_{-\infty}^{+\infty}d\phi_1d\phi_2 \int_{-\infty}^{\phi_1}d\varphi_1\int_{-\infty}^{\phi_2}d\varphi_2 e^{i \Phi_1(\varphi_1) +i \Phi_{21}(\phi_1)-i \Phi_2(\varphi_2) -i \Phi_{12}(\phi_2)}\times\nonumber\\
	&\left(\frac{\varepsilon}{m}\bm{\theta}_{pk_1}-\xi\bm{A}(\varphi_1)\right)\cdot\left(\frac{\varepsilon}{m}\bm{\theta}_{pk_1}-\xi\bm{A}(\phi_2)\right) \left(\frac{\varepsilon}{m}\bm{\theta}_{pk_2}-\xi\bm{A}(\phi_1)\right)\cdot\left(\frac{\varepsilon}{m}\bm{\theta}_{pk_2}-\xi\bm{A}(\varphi_2)\right)\,,\nonumber
\end{align}

The terms $A_{11}$, $A_{22}$, and $A_{12}$  correspond to $|S_1|^2$, $|S_2|^2$, and $S_1 S^*_2$, respectively. Here we also assume that the laser field has linear polarization, directed along $\bm e_x$. The summation over the photon polarizations is performed using
\begin{equation}\label{pol_sum}
	\pi_{p}e_i=-(\varepsilon\bm{\theta}_{pk_i}-e\bm{A})\cdot\bm{e_i}\,,\quad \sum_\lambda e_\lambda^ie_\lambda^j=\delta_\perp^{ij}\,,
\end{equation}
where $\delta_\perp^{ij}$ is a Kronecker delta in the $xy$-plane. The soft photon approximation for double Compton scattering is valid when both photon energies satisfy $\omega_{1,2} \ll \varepsilon$, similar to the single Compton case. 

The strength of this approach is that it also yields a compact expression for the case $\varepsilon\sim\omega_1$ and $\omega_2\ll\varepsilon$. Starting from the matrix element~\eqref{eq:DC1}, and using the formula for nonlinear Compton scattering~\eqref{eq:res_HE_exact}, we can obtain the probability of two-photon emission $\dfrac{d W^{(1)}}{d^3 k_1 d^3 k_2}$ in this case:

\begin{align}\label{eq:W1}
	&\dfrac{d W^{(1)}}{d^3 k_1 d^3 k_2}=\frac{\alpha^2 m^4}{(4\pi \varepsilon)^4\omega_1\omega_2} \left(A_{11}+A_{22}+2\Re A_{12} \right) \,, \\
	A_{11}&=\int_{-\infty}^{+\infty}d\phi_1d\phi_2 \int_{-\infty}^{\phi_1}d\varphi_1\int_{-\infty}^{\phi_2}d\varphi_2 e^{i \Phi_1(\varphi_1) +i \Phi_{21}(\phi_1)-i \Phi_1(\varphi_2) -i \Phi_{21}(\phi_2)}\times\nonumber\\
	&\left[g_1+g_2\left(\frac{\varepsilon}{m}\bm{\theta}_{pk_1}-\xi\bm{A}(\varphi_1)\right)\cdot\left(\frac{\varepsilon}{m}\bm{\theta}_{pk_1}-\xi\bm{A}(\varphi_2)\right)\right] \left(\frac{\varepsilon}{m}\bm{\theta}_{pk_2}-\xi\bm{A}(\phi_1)\right)\cdot\left(\frac{\varepsilon}{m}\bm{\theta}_{pk_2}-\xi\bm{A}(\phi_2)\right)\,,\nonumber\\
	A_{22}&=\int_{-\infty}^{+\infty}d\phi_1d\phi_2 \int_{-\infty}^{\phi_1}d\varphi_1\int_{-\infty}^{\phi_2}d\varphi_2 e^{i \Phi_2(\varphi_1) +i \Phi_{12}(\phi_1)-i \Phi_2(\varphi_2) -i \Phi_{12}(\phi_2)}\times\nonumber\\
	&\left[g_1+g_2\left(\frac{\varepsilon}{m}\bm{\theta}_{pk_1}-\xi\bm{A}(\phi_1)\right)\cdot\left(\frac{\varepsilon}{m}\bm{\theta}_{pk_1}-\xi\bm{A}(\phi_2)\right)\right] \left(\frac{\varepsilon}{m}\bm{\theta}_{pk_2}-\xi\bm{A}(\varphi_1)\right)\cdot\left(\frac{\varepsilon}{m}\bm{\theta}_{pk_2}-\xi\bm{A}(\varphi_2)\right)\,,\nonumber\\
	A_{12}&=\int_{-\infty}^{+\infty}d\phi_1d\phi_2 \int_{-\infty}^{\phi_1}d\varphi_1\int_{-\infty}^{\phi_2}d\varphi_2 e^{i \Phi_1(\varphi_1) +i \Phi_{21}(\phi_1)-i \Phi_2(\varphi_2) -i \Phi_{12}(\phi_2)}\times\nonumber\\
	&\left[g_1+g_2\left(\frac{\varepsilon}{m}\bm{\theta}_{pk_1}-\xi\bm{A}(\varphi_1)\right)\cdot\left(\frac{\varepsilon}{m}\bm{\theta}_{pk_1}-\xi\bm{A}(\phi_2)\right)\right] \left(\frac{\varepsilon}{m}\bm{\theta}_{pk_2}-\xi\bm{A}(\phi_1)\right)\cdot\left(\frac{\varepsilon}{m}\bm{\theta}_{pk_2}-\xi\bm{A}(\varphi_2)\right)\,,\nonumber\\
	&g_1=\frac{\omega_1^2}{2(\varepsilon-\omega_1)^2}\,,\quad g_2=\frac{\varepsilon^2+(\varepsilon-\omega_1)^2}{2(\varepsilon-\omega_1)^2}\,.\nonumber
\end{align}
This expression is more general than the double soft photon result and reduces to it in the limit $\omega_1 \ll \varepsilon$. Note that, while the probability of double Compton scattering Eqs.~\eqref{eq:W} and \eqref{eq:W1} is written in terms of four-dimensional integrals, it can be expressed as a sum of products of the two-dimensional integrals $f_{ij}$ and $g_{ij}$, which significantly simplifies the numerical evaluation. This form was chosen for the sake of brevity and transparency of the derivation, while the representation in terms of $f_{ij}$ and $g_{ij}$ is more convenient for actual computations.

Note that the integrands of the functions $f_{ij}$ and $g_{ij}$ factorize with respect to the integration variables $\phi$ and $\varphi$. It enables us to perform  the numerical computation of these functions with $O(N_{\text{grid}})$ complexity, as explained in Sec.~\ref{sec:num2}.

It is worth noting that in the present kinematics the integrals over the photon angles are Gaussian and can be evaluated analytically. However, in this case the integration variables $\phi_1$, $\phi_2$, $\phi'_1$, $\phi'_2$ become coupled, and the integral no longer factorizes into a product of one-dimensional integrals. Consequently, to compute the angle-integrated spectrum, one would need to evaluate the full four-dimensional integral directly with a sufficiently fine grid. This approach is likely to be more computationally demanding than computing the one-dimensional integrals for the functions $f_{ij}$ and $g_{ij}$ followed by Monte Carlo integration over the photon angles. The situation is further exacerbated for larger $N$, for example, for $N=3$, where the direct integration would require a six-dimensional integral.

\begin{figure}
	\centering
	\includegraphics[width=0.8\linewidth]{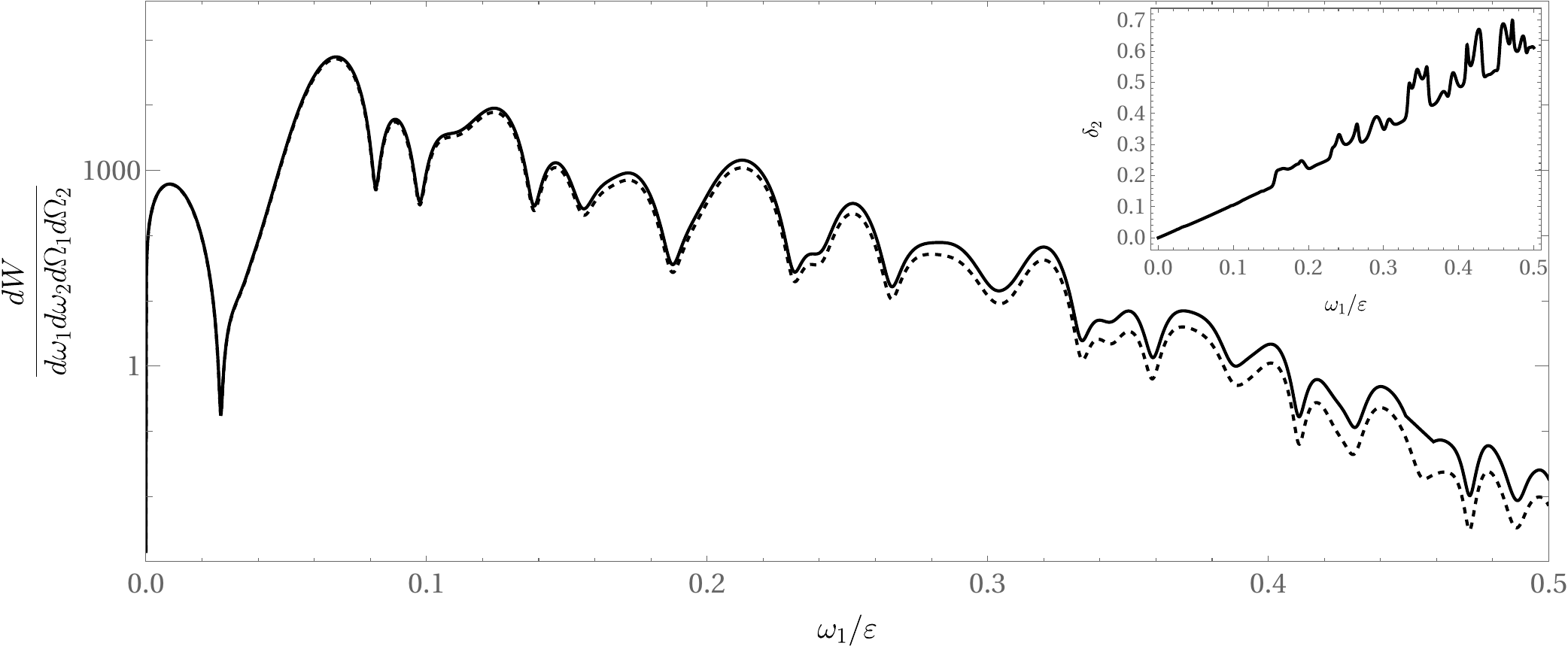}
	\caption{The probability of double nonlinear Compton scattering $\frac{d W}{d\omega_1d\omega_2 d\Omega_1d\Omega_2}$ as a function of  $\omega_1/\varepsilon$, the photon angles are $\theta_{pk_1x}=\frac{m}{4\varepsilon}$, $\theta_{pk_1y}=\frac{m}{2\varepsilon}$, $\theta_{pk_2x}=\frac{m}{3\varepsilon}$, $\theta_{pk_2y}=\frac{m}{\pi\varepsilon}$, and $\omega_2=\varepsilon/100$. The details of numerical computation are discussed in the text. The black line corresponds to  $d W^{(1)}$ and the dashed line corresponds to  $d W$. Inset: the relative difference $\delta_2$ between $d W^{(1)}$ and $d W$ as a function of  $\omega_1/\varepsilon$.}
	\label{pic:DCompton}
\end{figure}

To illustrate our analytical results, we perform numerical computations of the double nonlinear Compton scattering probability. The calculations use the same kinematic configuration as in Sec.~\ref{sec:SCS}.
In Fig.~\ref{pic:DCompton} we present the probabilities of double nonlinear Compton scattering $\frac{d W}{d\omega_1d\omega_2 d\Omega_1d\Omega_2}$ and $\frac{d W^{(1)}}{d\omega_1d\omega_2 d\Omega_1d\Omega_2}$ as a function of $\omega_1/\varepsilon$. The relative difference $\delta_2$ between $dW^{(1)}$ and $dW$ is approximately linear, with several additional peaks superimposed. The origin of these peaks is related to the fact that the functions $f_{ij}$ and $g_{ij}$ have different resonant frequency conditions. This result demonstrates that the accuracy of the soft photon approximation is  $\mathcal{O}\left(\frac{\omega_1}{\varepsilon}\right)$ the same as in single nonlinear Compton scattering. We note that, unlike the single Compton case, the soft photon approximation for double Compton scattering remains valid at zero emission angles, as the differential probability depends on multiple functions $f_{ij}$ and $g_{ij}$ rather than a single integral.

\begin{figure}
	\centering
	\includegraphics[width=0.6\linewidth]{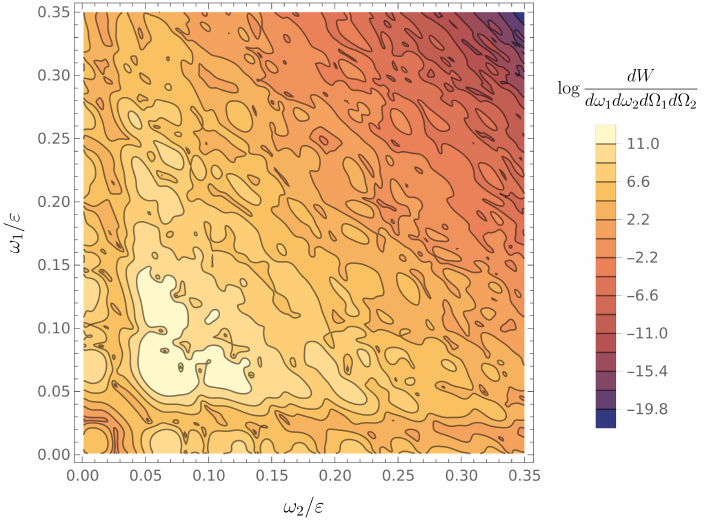}
	\caption{(color online). The logarithm of the probability of double nonlinear Compton scattering $\log\frac{d W}{d\omega_1d\omega_2 d\Omega_1d\Omega_2}$ as a function of  $\omega_1/\varepsilon$ and $\omega_2/\varepsilon$. The photon angles are $\theta_{pk_1x}=\frac{m}{4\varepsilon}$, $\theta_{pk_1y}=\frac{m}{2\varepsilon}$, $\theta_{pk_2x}=\frac{m}{3\varepsilon}$, $\theta_{pk_2y}=\frac{m}{\pi\varepsilon}$. The details of numerical computation are discussed in the text.}
	\label{pic:CountorDCompton}
\end{figure}

In Fig.~\ref{pic:CountorDCompton} a contour plot of the logarithm of the differential probability $\log\frac{d W}{d\omega_1d\omega_2 d\Omega_1d\Omega_2}$ in the $\omega_1/\varepsilon$ -- $\omega_2/\varepsilon$ plane for some specific angles is presented. It shows the nontrivial and highly varying dependence of this probability. This probability has been studied in detail in Refs.~\cite{mackenroth2013nonlinear, seipt2012two}.

\begin{figure}
	\centering
	\includegraphics[width=0.65\linewidth]{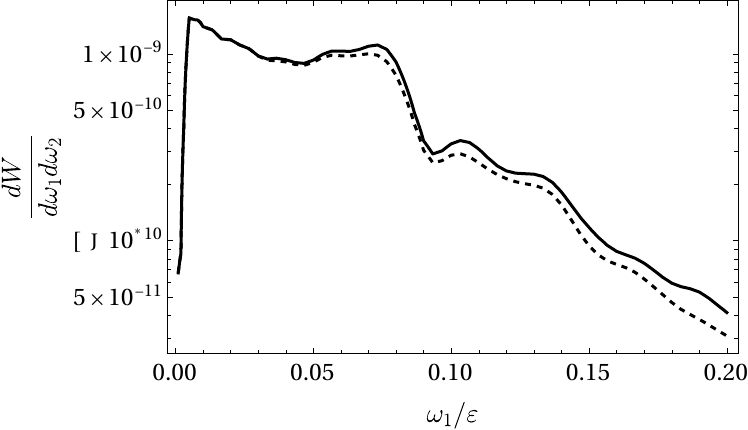}
	\caption{The probability of double nonlinear Compton scattering integrated over the photon angles as a function of  $\omega_1/\varepsilon$, $\omega_2=\varepsilon/100$. The details of numerical computation are discussed in the text. The black line corresponds to $d W^{(1)}$ (exact in $\omega_1$, soft photon approximation in $\omega_2$) and the dashed line corresponds to  $d W$ (soft photon approximation for both photons).}
	\label{pic:DCompton1}
\end{figure}

\begin{figure}
	\centering
	\includegraphics[width=0.6\linewidth]{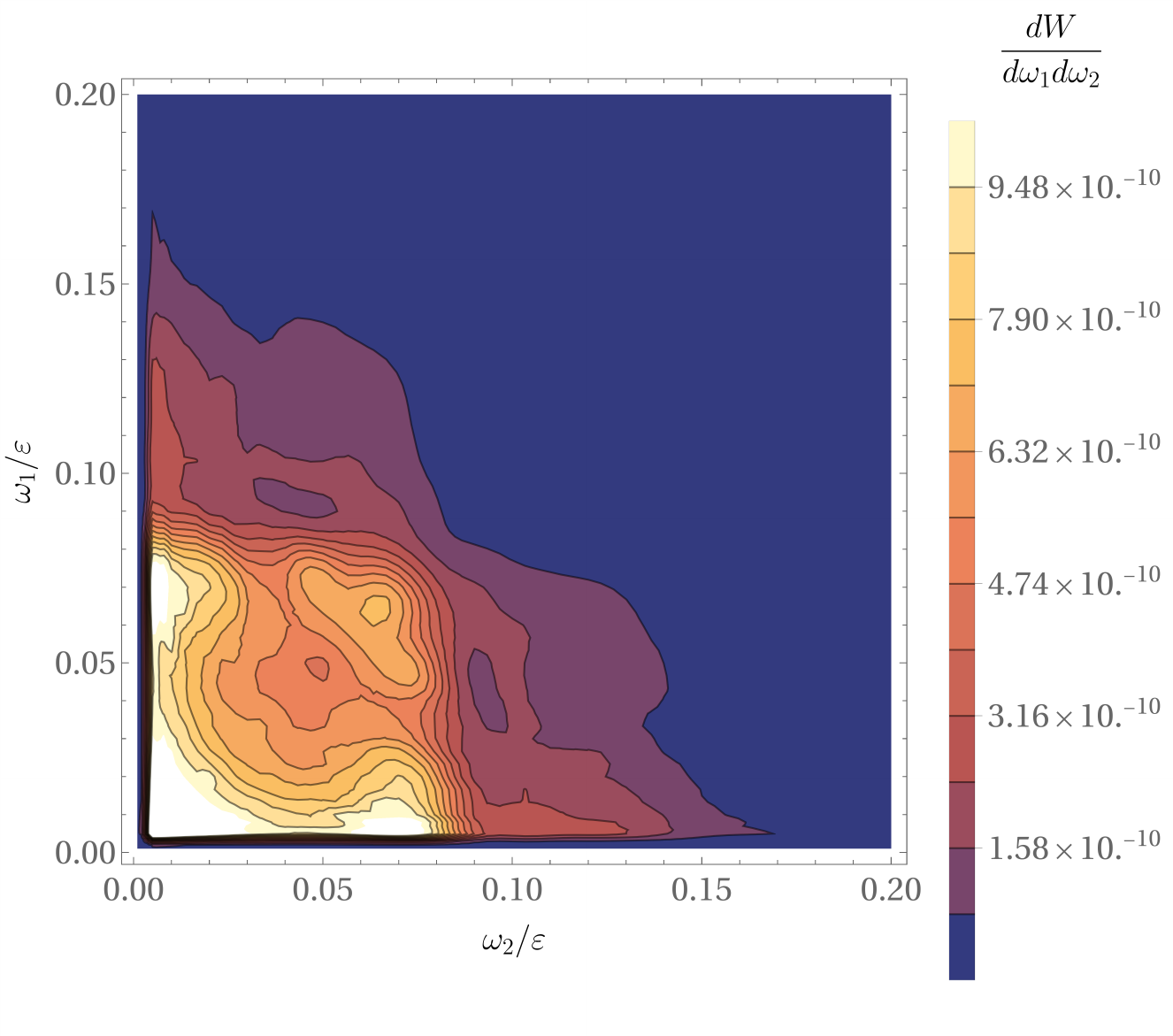}
	\caption{(color online). The angle-integrated probability of double nonlinear Compton scattering $\frac{d W}{d\omega_1d\omega_2}$ as a function of  $\omega_1/\varepsilon$ and $\omega_2/\varepsilon$. The details of numerical computation are discussed in the text.}
	\label{pic:CountorDCompton1}
\end{figure}

The probabilities integrated over the photon angles $\frac{d W}{d\omega_1d\omega_2}$ and $\frac{d W^{(1)}}{d\omega_1d\omega_2}$ are presented in Fig.~\ref{pic:DCompton1}. It is seen from Fig.~\ref{pic:DCompton1} that the accuracy of the soft photon approximation is $\mathcal{O}\left(\omega/\varepsilon\right)$. The relative difference reaches $30\%$ between $\frac{d W^{(1)}}{d\omega_1d\omega_2}$ and $\frac{d W}{d\omega_1d\omega_2}$ at $\frac{\omega_1}{\varepsilon}=0.2$ and $\frac{\omega_2}{\varepsilon}=0.01$, as shown in Fig.~\ref{pic:DCompton1}. In Fig.~\ref{pic:CountorDCompton1} a contour plot of the angle-integrated probability $\frac{d W}{d\omega_1d\omega_2 }$ in the $\omega_1/\varepsilon$ -- $\omega_2/\varepsilon$ plane is presented. It shows the nontrivial dependence of this probability but it is smoother in comparison 
with  Fig.~\ref{pic:CountorDCompton}.

\begin{figure}
	\centering
	\includegraphics[width=0.7\linewidth]{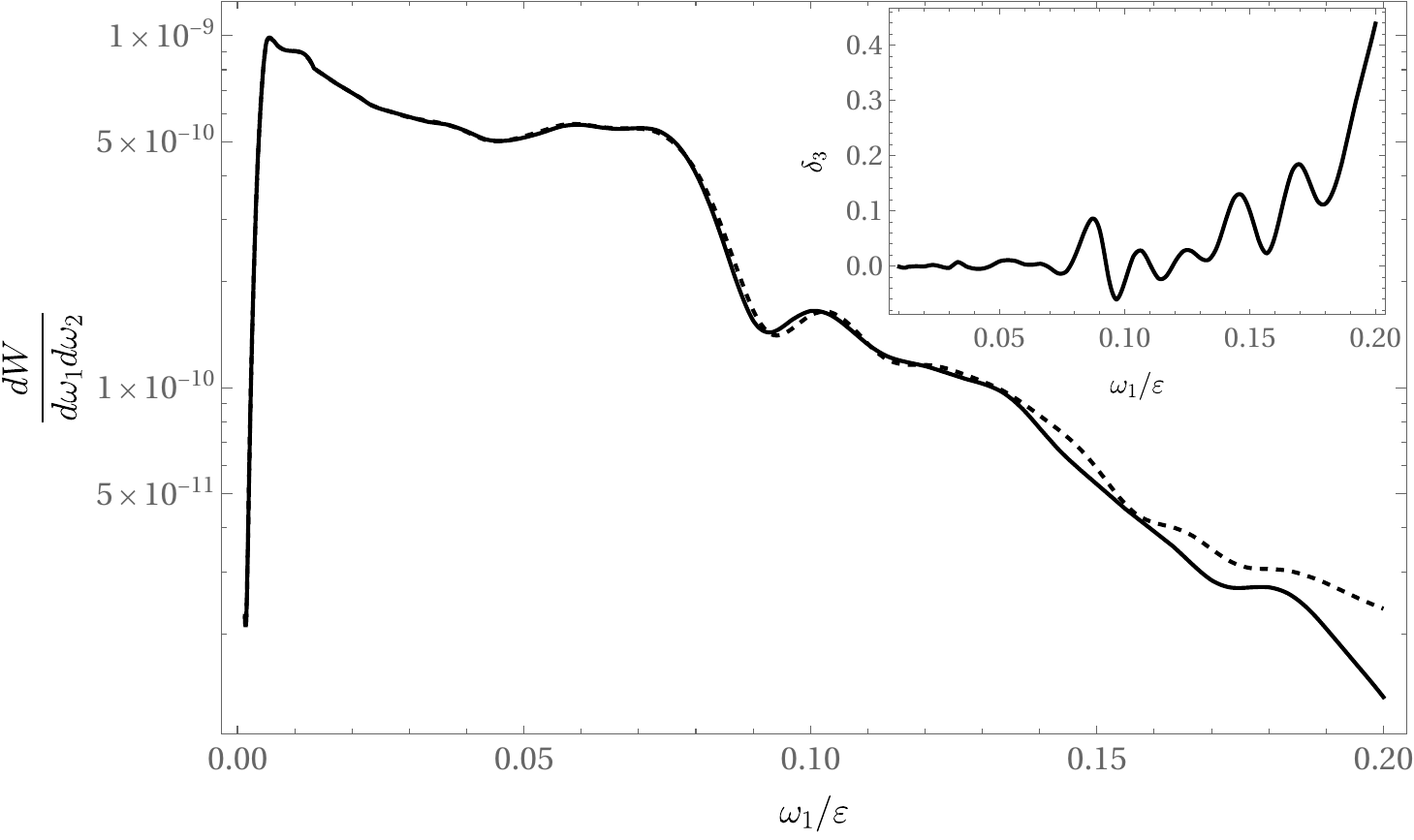}
	\caption{The probability of double nonlinear Compton scattering integrated over the photon angles as a function of $\omega_1/\varepsilon$, $\omega_2=\varepsilon/25$. The details of numerical computation are discussed in the text. The black line corresponds to $d W$ (soft photon result) and the dashed line corresponds to  $d W_c$ (the cascade contribution~\eqref{eq:Wc} in the soft-photon approximation). Inset: the relative difference $\delta_3$ between $d W$ and $d W_c$ as a function of $\omega_1/\varepsilon$.}
	\label{pic:DCompton2}
\end{figure}

In order to simulate the interaction of electrons with a laser pulse, PIC codes are widely used. These codes primarily rely on the probabilities of single nonlinear Compton scattering and Breit–Wheeler pair production to simulate a QED cascade in various field configurations. In such codes, the emission of two photons is typically modeled as two independent single nonlinear Compton scattering events. Thus, it is interesting to discuss interference effects in nonlinear double Compton scattering and compare our results with the cascade contribution, where the two photons are emitted independently. The interference term cannot be simply dropped from the differential probability~\eqref{eq:W}, as doing so would violate gauge invariance. In order to construct the cascade contribution to the probability integrated over the photon angles, we use the result of~\cite{PhysRevD.99.096018}. Specifically, the cascade contribution $\frac{dW_c}{d\omega_1d\omega_2}$ to the angle-integrated probability of double nonlinear Compton scattering is given by:
\begin{align}\label{eq:Wc}
	\dfrac{d W_c}{d\omega_1d\omega_2}=\frac{1}{2}\left(\frac{dW(p)}{d\omega_1}\frac{dW({\cal P}(p,-k_1))}{d\omega_2}+1\leftrightarrow 2\right)\,,
\end{align}
where $\frac{dW(q)}{d\omega}$ is the probability of single nonlinear Compton scattering of an electron with initial momentum $q$ in the soft photon approximation.
The comparison of $\frac{dW_c}{d\omega_1d\omega_2}$ and $\frac{dW}{d\omega_1d\omega_2}$ as a function of $\omega_1/\varepsilon$ at $\omega_2/\varepsilon=0.04$ is shown in Fig.~\ref{pic:DCompton2}. In the region  $\omega_1<0.08\varepsilon$ these two quantities have a small relative difference $\delta_3<1$\%, while in the intermediate region $0.08\varepsilon<\omega_1<0.2\varepsilon$ the difference becomes significant and grows with increasing photon energy. At $\omega_1=0.2\varepsilon$ the relative difference reaches $\delta_3=44$\%.

Note that the result of~\cite{PhysRevD.99.096018} for the cascade contribution~\eqref{eq:Wc} is valid only for the probability integrated over photon angles. For the differential probability there is no analogous result. This is related to the fact that the cascade contribution comes from the region where the intermediate electron is on the mass shell. For arbitrary angles of the emitted photons the intermediate electron is off the mass shell. Thus, the fully differential probability of nonlinear double Compton scattering cannot be factorized into a product of single-Compton probabilities and requires the coherent treatment developed in this work.

\subsection{$N$ Compton scattering}\label{sec:NCS}
In the present subsection we discuss the amplitude of $N$-photon emission in the soft photon approximation. Multiphoton emission was studied in \cite{PhysRevD.99.096018}, where the dominant cascade contribution for long laser pulses was discussed. In that approach, the probability of multiple photon emission is obtained as an incoherent product of single-photon emission probabilities, where the photons are emitted independently. However, a systematic analytical expression for the full $N$-photon differential probability that accounts for interference between different amplitudes was not provided. This is in contrast to the coherent $N$-photon emission considered in the present work, where all $N$ photons are emitted in a single quantum process with interference between different emission orderings. Note that the cascade contribution is automatically captured within our 
approach.

It is well known that in the classical theory of radiation the photons are emitted independently. This fact corresponds to the factorization of the amplitude for multiple photon emission, as was shown in~\cite{ilderton2013scattering,Krachkov24}. The soft photon approximation takes into account the change of the electron momentum after the soft photon emission in the phase, which corresponds to replacing $p$ with ${\cal P}(p,-k_{i})$. This replacement breaks the factorization, as the intermediate electron momenta in different amplitudes are different. This is a key difference between the soft photon and classical approximations: the former captures the quantum recoil effect, while the latter does not.

\begin{figure}[htbp]
	\centering
	\begin{tikzpicture}
		\begin{feynman}
			\vertex (i);
			\vertex [right=1.5cm of i] (v1);
			\vertex [right=6.5cm of v1] (v2);
			\vertex [right=7.5cm of v2] (v3);
			\vertex [right=1.5cm of v3] (o);			
			\vertex [above right=1.2cm and 0.8cm of v1] (p1);
			\vertex [above right=1.2cm and 0.8cm of v2] (p2);
			\vertex [above right=1.2cm and 0.8cm of v3] (p3);			
			\coordinate [left=3.3cm of v2] (mid1);
			\coordinate [left=3.3cm of v3] (mid2);			
			\draw [fermion] (i) -- node[below=0.15cm] {\(p\)} (v1);
			\draw [fermion] (v1) -- node[below=0.15cm] 
			{\({\cal P}(p,-k_{\sigma(1)})\,\,\,\quad\cdots\quad\,\,\,{\cal P}(p,-\sum\limits_{j<i} k_{\sigma(j)})\)} (v2);
			\draw [fermion] (v2) -- node[below=0.15cm] 
			{\({\cal P}(p,-\sum\limits_{j\leq i} k_{\sigma(j)})\,\,\,\quad\cdots\quad\,\,\,{\cal P}(p,-\sum\limits_{j\leq N} k_{\sigma(j)})\)} (v3);
			\draw [fermion] (v3) -- node[below=0.15cm] {\(p'\)} (o);			
			\draw [photon] (v1) -- node[above=0.1cm, sloped] {\(k_{\sigma(1)}\)} (p1);
			\draw [photon] (v2) -- node[above=0.1cm, sloped] {\(k_{\sigma(i)}\)} (p2);
			\draw [photon] (v3) -- node[above=0.1cm, sloped] {\(k_{\sigma(N)}\)} (p3);			
			\node[below=0.3cm of v1] {\(\varphi_1\)};
			\node[below=0.3cm of v2] {\(\varphi_i\)};
			\node[below=0.3cm of v3] {\(\varphi_N\)};
			\node[above=0.5cm of mid1] {\Huge\(\cdots\cdots\)};
			\node[above=0.5cm of mid2] {\Huge\(\cdots\cdots\)};			
		\end{feynman}
	\end{tikzpicture}
	\caption{The Feynman diagram for  $N$-photon nonlinear Compton scattering which corresponds to the emission order $k_{\sigma(1)},\,k_{\sigma(2)}\,\ldots,\,k_{\sigma(N)}$ for permutation $\sigma$. Here, we also depict the phase of each interaction point and the momenta of the intermediate electron.}
	\label{fig:Nphotons}
\end{figure}
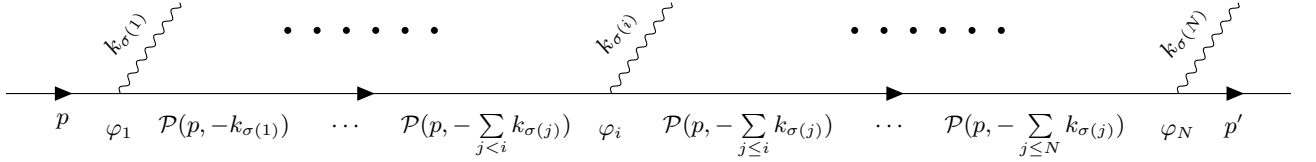

In the present subsection we discuss the following process. The electron with initial momentum $p$ emits $N$ photons with momenta $k_i$, \( i \in \{1, 2, \ldots, N\} \).  In order to enumerate all diagrams, we use the permutation group  $\mathcal{S}_N$. For a given permutation $\sigma$, the electron emits photons in the following order: $k_{\sigma(1)},\,k_{\sigma(2)}\,\ldots,\,k_{\sigma(N)}$. The Feynman diagram for this amplitude $S_\sigma$ is shown in Fig.~\ref{fig:Nphotons}.

Following Eq.~\eqref{eq:2photon}, the amplitude $S_\sigma$ is: 
\begin{align}\label{eq:SN}
	&S_\sigma=(-i)^{N-1}2p_-(2\pi)^3 \delta\left(\bm p_\perp-\bm p'_\perp-\sum_{i\leq N}\bm k_{i\perp}\right) \delta\left(p_-- p'_-- \sum_{i\leq N} k_{i-}\right)\times \\
	&\prod_{i\leq N} \left[\int_{-\infty}^{\varphi_{i+1}}\frac{d\varphi_i}{p_{-}}e^{i \Phi_{\sigma,i}(\varphi_i)}(\pi_{p}(\varphi_i)e_{\sigma(i)}^{*})\right] 
	\,,\nonumber\\
	&\Phi_{\sigma,i}(\varphi_i)=\int_{0}^{\varphi_i}d\varphi'_i\frac{(\pi_{{\cal P}(p,-\sum_{j<i} k_{\sigma(j)})}(\varphi_i')k_{\sigma(i)})}{{\cal P}_-(p,-\sum_{j\leq i} k_{\sigma(j)})}\,.
\end{align}
The integration limits $\varphi_i < \varphi_{i+1}$ reflect the time ordering of photon emissions: in the amplitude $S_\sigma$, the photons are emitted in the order $k_{\sigma(1)}, k_{\sigma(2)}, \ldots, k_{\sigma(N)}$, with $\varphi_1 < \varphi_2 < \ldots < \varphi_N<\varphi_{N+1}=+\infty$. The total amplitude is a sum of $S_\sigma$ for all permutations:
\begin{equation}
	S=\sum_{\sigma\in \mathcal{S}_N} S_\sigma\,.
\end{equation}
As shown in our previous work~\cite{Krachkov24}, the sum over all permutations $\sigma \in \mathcal{S}_N$ of the integrals over the simplices $\varphi_1 < \varphi_2 < \ldots < \varphi_N$, neglecting $k_i$ in comparison with $p$ in the phase, yields the product of independent integrals from $-\infty$ to $+\infty$, which leads to the factorization of the amplitude in the classical limit.

Note that for an ultra-relativistic electron in a head-on collision with a laser pulse, the photons are emitted at small angles. 
\begin{equation}\label{eq:phaseN}
	\frac{(\pi_{{\cal P}(p,-\sum_{j<i} k_{\sigma(j)})}(\varphi_i')k_{\sigma(i)})}{{\cal P}_-(p,-\sum_{j\leq i} k_{\sigma(j)})}=
	\frac{\omega_{\sigma(i)}\left[m^2+\left(\left(\varepsilon-\sum_{j<i}\omega_{\sigma(j)} \right)\bm\theta_{pk_{\sigma(i)}}+\sum_{j<i}\omega_{\sigma(j)}\bm\theta_{pk_{\sigma(j)}}-e\bm A\right)^2\right]}{4\left(\varepsilon-\sum_{j<i}\omega_{\sigma(j)} \right)\left(\varepsilon-\sum_{j\leq i}\omega_{\sigma(j)} \right)}\,,
\end{equation}
where $\omega_i$ is the frequency of the photon with  momentum $k_i$, $\bm \theta_{pk_i}$ is a two-dimensional angle between the electron momentum and the momentum $k_i$: $\bm\theta_{pk_i}=\frac{\bm p_\perp}{\varepsilon_p}-\frac{\bm k_{i\perp}}{\omega_i}$.
Using Eq.~\eqref{pol_sum}, it is easy to obtain the probability of $N$-photon emission in analogy with Eq.~\eqref{eq:W}

\begin{align}\label{eq:WN}
	&dW_N =\left(\frac{\alpha}{16\pi^2\varepsilon^2}\right)^N
	\sum_{\sigma,\sigma' \in \mathcal{S}_N}
	\prod_{i,k \le N}\frac{d^3k_i}{\omega_i}
	\int_{-\infty}^{\varphi_{i+1}} d\varphi_i e^{i \Phi_{\sigma,i}(\varphi_i)}
	\int_{-\infty}^{\varphi'_{k+1}} d\varphi'_k e^{-i \Phi_{\sigma',k}(\varphi'_k)}\times\nonumber\\
	&\left(\varepsilon\bm{\theta}_{pk_{\sigma(i)}}-e\bm{A}(\varphi_i)\right)^{n_{\sigma(i)}}\left(\varepsilon\bm{\theta}_{pk_{\sigma'(k)}}-e\bm{A}(\varphi_k)\right)^{n_{\sigma'(k)}}\,,
\end{align}
where the phases are determined by Eqs.~\eqref{eq:phaseN}. Here the indices $n_{\sigma(i)}$ and $n_{\sigma'(k)}$ indicate which photon corresponds to each vector. The summation over photon polarizations, given by Eq.~\eqref{pol_sum}, contracts the vectors that correspond to the same photon, i.e., when $\sigma(i) = \sigma'(k)$. For example, for $N=2$ and permutations $\sigma=(1,2)$ and $\sigma'=(2,1)$, the contraction structure is:
\begin{equation}
	\left(\varepsilon\bm{\theta}_{pk_1}-e\bm{A}(\varphi_1)\right) \cdot \left(\varepsilon\bm{\theta}_{pk_1}-e\bm{A}(\varphi'_2)\right) \times \left(\varepsilon\bm{\theta}_{pk_2}-e\bm{A}(\varphi_2)\right) \cdot \left(\varepsilon\bm{\theta}_{pk_2}-e\bm{A}(\varphi'_1)\right)\,,
\end{equation}
where the first vector from $S_\sigma$ (corresponding to photon 1) is contracted with the second vector from $S_{\sigma'}$ (also corresponding to photon 1), and vice versa.

This result for $N=2$ coincides with Eq.~\eqref{eq:W}. Note that for a more general case (arbitrary electron energy and arbitrary angles between electron momenta and laser bunch direction) we can also obtain the analogous result using the matrix element Eq.~\eqref{eq:SN} and the photon polarization summation rule:
\begin{equation}
	\sum_\lambda e_\lambda^\mu e_\lambda^\nu=-g^{\mu\nu}\,,\quad \sum_\lambda(\pi_{p}(\varphi_1)e_\lambda)(\pi_{p}(\varphi_2)e^*_\lambda)=-\left(m^2-\frac{e^2}{2}(A(\varphi_1)-A(\varphi_2))^2\right)\,.
\end{equation}
We will not write it here due to the straightforward modification of Eq.~\eqref{eq:WN}.
The soft photon approximation for $N$-photon emission is valid when all photon energies satisfy $\omega_i \ll \varepsilon$. However, for large $N$, the total emitted energy should also be much smaller than $\varepsilon$: $\sum_{i=1}^N \omega_i\ll\varepsilon$. The accuracy of Eq.~\eqref{eq:WN} is $\mathcal{O}\left(\frac{\sum_{i=1}^N \omega_i}{\varepsilon}\right)$.

The power of the soft photon approximation is that it provides a compact analytical expression for the $N$-photon emission probability, which would be extremely difficult to obtain by direct calculation. The result Eq.~\eqref{eq:WN} can be evaluated numerically for any $N$ using standard Monte Carlo integration methods. The case $N > 2$ is particularly relevant for the description of cascade processes in strong-field QED, where an electron emits multiple photons in a sequence of hard and soft emissions. The soft photon approximation provides a tractable framework for studying such processes.

Note that the numerical evaluation of the differential probability for $N$-photon emission,  Eq.~\eqref{eq:WN},  can be performed with $\mathcal{O}(N_{\text{grid}})$ complexity using the same approach as for double Compton scattering, since the integrands factorize with respect to the integration variables.

\section{Numerical computations}\label{sec:NR}
In this section we discuss the  methods used to obtain numerical results. The  computation of the matrix element for a QED process in a plane wave background  requires the evaluation of highly oscillatory integrals. One of the promising methods for such computations is the Levin-type method \cite{levin1982procedures} and its modifications \cite{levin1996fast, iserles2004quadrature}. These methods work well for the one-dimensional case.  However, the extension of Levin-type methods to multidimensional integrals remains an open problem.

In the present paper we adopt a different approach. We use the simplest trapezoidal method for numerical computation. The advantage of the trapezoidal method is that it converges quickly for oscillating functions, since the error over the period is eliminated. Furthermore, the multidimensional integrals for double and $N$-photon Compton scattering can be reduced to one-dimensional form with $O(N)$ complexity, making sophisticated multidimensional methods unnecessary and favoring the simple, fast trapezoidal rule. 
The trapezoidal rule is particularly well-suited for smooth oscillatory integrands. For periodic functions, the trapezoidal rule exhibits exponential convergence, a property known from the Euler-Maclaurin formula~\cite{trefethen2014exponentially}. For highly oscillatory integrals, the error contributions from successive oscillation periods tend to cancel, leading to rapid convergence even when the integrand oscillates many times over the integration domain. This makes the trapezoidal rule more efficient than higher-order methods such as Simpson's rule for oscillatory integrands.

The specific properties of the laser pulse are given in Sec.~\ref{sec:SCS} after Eq.~\eqref{eq:res_HE_exact}, but we repeat them here for completeness. We consider electrons with initial energy $\varepsilon=10^4 m$ in head-on collisions with a linearly polarized laser pulse. Calculations are performed for $\xi=1$ and a pulse shape  $\bm A(\phi)= \frac{m \xi}{|e|} \cos[\omega_0\phi] g(\phi)\bm e_x$, with the envelope function $g(\phi)=\cos^2(\pi \omega_0 \phi/2\tau)$ for $-\tau\leq\omega_0 \phi\leq \tau$ and zero otherwise, with a dimensionless pulse length $\tau=20$ and $\omega_0=1.55$ eV. The electron charge is negative $e=-|e|$. We set $m=1$, thus all numerical quantities are presented in these units.

\subsection{Numerical evaluation of single nonlinear Compton scattering}\label{sec:num1}
For the nonlinear single Compton scattering, we compute the phase difference during the laser pulse duration $\Delta\Phi=\Phi(\tau/\omega_0)-\Phi(-\tau/\omega_0)$, where $\Phi(\varphi)$ is the phase in Eq.~\eqref{eq:res_HE}. The number of points for the one-dimensional grid is $N_{\text{points}}=\frac{N_0}{10}\max\left(\frac{\Delta\Phi}{2\pi},10\right)$. It means that for one period we have at least $\frac{N_0}{10}$ points. For practical computations we use $N_0=1000$. We compute the functions $f_{1,2}$ from Eq.~\eqref{eq:res_HE} directly by the trapezoidal method, since they are nonzero only in a finite segment $-\tau/\omega_0\leq \phi\leq \tau/\omega_0$. The function $f_0$ involves integration over infinite limits, which requires careful treatment. Following the standard prescription in quantum field theory, we understand the infinite limits in the sense of adiabatic switching of the interaction: the interaction is switched on exponentially in the distant past and switched off in the distant future. This is implemented by introducing regularization parameters $(1\pm i0)$ in the phase, which ensure exponential convergence of the integrals at infinity.

Explicitly, we decompose $f_0$ as:
\begin{align}\label{eq:f0_decomp}
	&f_0=\int_{-\tau/\omega_0}^{+\tau/\omega_0} d\phi\, e^{i\Phi(\phi)}+ e^{i\Phi(\tau/\omega_0)}\int_{0}^{+\infty} d\phi\, e^{i\frac{\omega \left[m^{2}+\varepsilon^2{\theta}_{pk}^2\right] \phi (1+i 0) }{4\varepsilon(\varepsilon-\omega)}} +\nonumber\\
	&e^{i\Phi(-\tau/\omega_0)}\int_{-\infty}^{0} d\phi\, e^{i\frac{\omega \left[m^{2}+\varepsilon^2{\theta}_{pk}^2\right] \phi (1-i 0) }{4\varepsilon(\varepsilon-\omega)}}\,,
\end{align}
where $\Phi(\phi)=\int_{0}^{\phi}\frac{\omega d\varphi }{4\varepsilon(\varepsilon-\omega)}\left[m^{2}+(\varepsilon\bm{\theta}_{pk}-e\bm{A}(\varphi))^{2}\right]$ is the phase. The integrals over the semi-infinite regions can be evaluated analytically, yielding:
\begin{equation}
	f_0 = \int_{-\tau/\omega_0}^{+\tau/\omega_0} d\phi\, e^{i\Phi(\phi)}-\frac{8\varepsilon(\varepsilon-\omega)}{\left[m^{2}+\varepsilon^2{\theta}_{pk}^2\right]\omega}\sin\left(\Phi(\tau/\omega_0)\right)\,,
\end{equation}
where we have used the symmetry $\Phi(-\phi)=-\Phi(\phi)$ of the phase for the symmetric laser pulse.

For fast evaluation we use the procedure \texttt{Compile}  in Wolfram Mathematica. Comparison with the built-in \texttt{NIntegrate} function shows that our implementation is approximately $10^3$ times faster for the required accuracy.

To verify the convergence of the numerical integration, we compute the functions $f_i$ with an increasing number of integration points and monitor the stability of the result. We find that for $N_0=1000$ (corresponding to 100 points per oscillation period), the relative change in $f_0$ when increasing  the number of points by a factor of $10$ is  of the order of $10^{-3}$ or less, which is sufficient for our purposes. For the functions $f_{1,2}$, the relative change is significantly smaller, of the order of $10^{-10}$. This is related to the fact that the integrands for  $f_{1,2}$ vanish at the boundaries, while that for $f_0$  is finite. Note that increasing the photon energy results in a decrease in the relative errors. In such cases, we increase the parameter $N_0$ to achieve relative errors of the order of  $10^{-3}$.  We also use the identity
\begin{align}
	&\int_{-\infty}^{+\infty} d e^{i\int_{0}^{\phi}\frac{\omega d\varphi }{4\varepsilon(\varepsilon-\omega)}\left[m^{2}+(\varepsilon\bm{\theta}_{pk}-e\bm{A}(\varphi))^{2}\right]}=f_0\left(1+\frac{\varepsilon^2\theta_{pk}^2}{m^2}\right)-2\xi\frac{\varepsilon\bm\theta_{pk}}{m}\cdot\bm f_1+\xi^2 f_2=0\,.\nonumber
\end{align}
for error estimation.

On a standard personal computer, the computational time is $5\times 10^{-4}$ s. Such a small computational time enables one to use Monte Carlo methods for numerical integration over the photon angles. We use the standard \texttt{NIntegrate} with the methods \texttt{AdaptiveMonteCarlo} or \texttt{QuasiMonteCarlo} for this purpose. 
The convergence of the Monte Carlo integration is verified by increasing the number of sampling points and monitoring the stability of the result. We find that $10^5$ points are sufficient to achieve a relative accuracy of $10^{-2}$ for the angular integration. 

We attach the \texttt{Wolfram Mathematica} notebook \texttt{Single Compton.nb} as an ancillary file. It contains the numerical computation of the functions $f_i$ and the differential spectra for the exact, soft photon, and classical results, given in Eqs.~\eqref{eq:res_HE} and \eqref{eq:res_HE_exact}.

\subsection{Numerical evaluation of nonlinear double Compton scattering}\label{sec:num2}

In this subsection we consider the numerical evaluation of the functions $f_{ij}$ and $g_{ij}$ from  Eq.~\eqref{eq:fgij}  arising in  the double Compton scattering probability. The functions $f_{ij}$  are defined as two-dimensional integrals:
\begin{equation}
	f_{i,j}=\int_{-\infty}^{+\infty}d\phi \int_{-\infty}^{\phi}d\varphi \left(\frac{e \bm A(\phi)}{m\xi}\right)^i\left(\frac{e \bm A(\varphi)}{m\xi}\right)^j  e^{i \Phi_1(\varphi) +i \Phi_{21}(\phi)}\,,
\end{equation}
where $\Phi_1(\varphi)$ and $\Phi_{21}(\phi)$ are the phases defined in Eqs.~\eqref{eq:fgij}. The functions $g_{ij}$ are obtained by the substitution $\Phi_1 \to \Phi_2$ and $\Phi_{21} \to \Phi_{12}$.

The key property that simplifies the numerical evaluation is the factorization of the integrand with respect to the integration variables $\phi$ and $\varphi$. The phases $\Phi_1(\varphi)$ and $\Phi_{21}(\phi)$ depend on different variables, therefore the integrand can be written as a product:
\begin{equation}
	f_{i,j} = \int_{-\infty}^{+\infty} d\phi \; h_i(\phi) \int_{-\infty}^{\phi} d\varphi \; g_j(\varphi)\,,
\end{equation}
where $h_i(\phi) = \left(\frac{e \bm A(\phi)}{m\xi}\right)^i e^{i\Phi_{21}(\phi)}$ and $g_j(\varphi) = \left(\frac{e \bm A(\varphi)}{m\xi}\right)^j e^{i\Phi_1(\varphi)}$.

Introducing the primitive function $G(\phi) = \int_{-\infty}^{\phi} d\varphi \; g_j(\varphi)$, we can rewrite the double integral as a single integral:
\begin{equation}
	f_{i,j} = \int_{-\infty}^{+\infty} d\phi \; h_i(\phi) \; G(\phi)\,.
\end{equation}

The function $G(\phi)$ is updated incrementally using the trapezoidal rule: $G(\phi_{n+1}) = G(\phi_n) + \frac{h}{2}[g_j(\phi_n) + g_j(\phi_{n+1})]$, where $h$ is the grid spacing. This reduces the computational complexity from $O(N_{\text{grid}}^2)$ for a direct two-dimensional integration to $O(N_{\text{grid}})$ for a single pass over the grid.

The laser pulse is finite: $\bm A(\phi) = 0$ for $|\phi| > \tau/\omega_0$. However, the phases $\Phi_1(\varphi)$ and $\Phi_{21}(\phi)$ continue to grow linearly outside the pulse region, since the electron propagates freely when the field is absent:
\begin{equation}
	\Phi_1(\phi) = \Phi_1(\pm\tau/\omega_0) + v_1 (\phi \mp \tau/\omega_0)\,, \quad \phi \gtrless \pm\tau/\omega_0\,,
\end{equation}
\begin{equation}
	\Phi_{21}(\phi) = \Phi_{21}(\pm\tau/\omega_0) + v_{21} (\phi \mp \tau/\omega_0)\,, \quad \phi \gtrless \pm\tau/\omega_0\,,
\end{equation}
where the asymptotic phase velocities are:
\begin{equation}
	v_1 = \frac{\omega_1(m^2+\varepsilon^2\bm\theta_{pk_1}^2)}{4\varepsilon(\varepsilon-\omega_1)}\,, \quad v_{21} = \frac{\omega_2[m^2+((\varepsilon-\omega_1)\bm\theta_{pk_2} + \omega_1\bm\theta_{pk_1})^2]}{4(\varepsilon-\omega_1)(\varepsilon-\omega_1-\omega_2)}\,.
\end{equation}

For $i \neq 0$ or $j \neq 0$, the factors $A(\phi)^i$ or $A(\varphi)^j$ vanish outside the pulse, and the integration is restricted to the finite region $[-\tau/\omega_0, \tau/\omega_0]$. However, for $i = 0$ or $j = 0$, the corresponding factors equal unity, and the integrand continues to oscillate outside the pulse. These ``tail'' contributions must be evaluated analytically, following the same regularization prescription as for the function $f_0$ in single Compton scattering (see Sec.~\ref{sec:num1}): the infinite integration limits are understood in the sense of adiabatic switching of the interaction, implemented by introducing regularization parameters $(1 \pm i0)$ in the phase.

We decompose the integration domain into three regions:
\begin{equation}
	\int_{-\infty}^{\infty} d\phi \int_{-\infty}^{\phi} d\varphi = \int_{-\infty}^{-\tau/\omega_0} d\phi \int_{-\infty}^{\phi} d\varphi + \int_{-\tau/\omega_0}^{\tau/\omega_0} d\phi \int_{-\infty}^{\phi} d\varphi + \int_{\tau/\omega_0}^{\infty} d\phi \int_{-\infty}^{\phi} d\varphi\,.
\end{equation}

The first region ($\phi < -\tau/\omega_0$, $\varphi < \phi$) contributes only when $i = j = 0$:
\begin{equation}
	\text{I} = \int_{-\infty}^{-\tau/\omega_0} d\phi \; e^{i[\Phi_{21}(-\tau/\omega_0) + v_{21}(\phi+\tau/\omega_0)]} \int_{-\infty}^{\phi} d\varphi \; e^{i[\Phi_1(-\tau/\omega_0) + v_1(\varphi+\tau/\omega_0)]}\,.
\end{equation}
The inner integral yields $\frac{e^{iv_1(\phi+\tau/\omega_0)}}{iv_1}$, and the outer integral gives:
\begin{equation}
	\text{I} = \frac{-e^{i[\Phi_1(-\tau/\omega_0) + \Phi_{21}(-\tau/\omega_0)]}}{v_1(v_{21}+v_1)}\,.
\end{equation}

The second region ($-\tau/\omega_0 \leq \phi \leq \tau/\omega_0$) is evaluated numerically. The initial value of $G(\phi)$ at $\phi = -\tau/\omega_0$ includes the contribution from $\varphi < -\tau/\omega_0$:
\begin{equation}
	G(-\tau/\omega_0) = \delta_{j,0}\int_{-\infty}^{-\tau/\omega_0} d\varphi \; e^{i\Phi_1(\varphi)} = \frac{e^{i\Phi_1(-\tau/\omega_0)}}{iv_1}\,.
\end{equation}

The third region ($\phi > \tau/\omega_0$) contains two contributions. The first arises from $\varphi < \tau/\omega_0$ and is nonzero when $i = 0$:
\begin{equation}
	\text{III}_a = \int_{\tau/\omega_0}^{\infty} d\phi \; e^{i[\Phi_{21}(\tau/\omega_0) + v_{21}(\phi-\tau/\omega_0)]} \cdot G(\tau/\omega_0) = \frac{i}{v_{21}} e^{i\Phi_{21}(\tau/\omega_0)} G(\tau/\omega_0)\,.
\end{equation}
The second contribution arises from $\varphi > \tau/\omega_0$ and is nonzero only when $i = j = 0$:
\begin{equation}
	\text{III}_b = \int_{\tau/\omega_0}^{\infty} d\phi \int_{\tau/\omega_0}^{\phi} d\varphi \; e^{i[\Phi_1(\tau/\omega_0) + v_1(\varphi-\tau/\omega_0)] + i[\Phi_{21}(\tau/\omega_0) + v_{21}(\phi-\tau/\omega_0)]} = \frac{-e^{i[\Phi_1(\tau/\omega_0) + \Phi_{21}(\tau/\omega_0)]}}{v_{21}(v_{21}+v_1)}\,.
\end{equation}

The total result is the sum of the numerical integral over the pulse region and the analytical tail contributions:
\begin{equation}
	f_{i,j} = \int_{-\tau/\omega_0}^{\tau/\omega_0} d\phi \; h_i(\phi) \; G(\phi) + \delta_{i0}\delta_{j0} \cdot \text{I} + \delta_{i0} \cdot \text{III}_a + \delta_{i0}\delta_{j0} \cdot \text{III}_b\,.
\end{equation}

The functions $g_{ij}$ are computed analogously by replacing $\Phi_1 \to \Phi_2$, $\Phi_{21} \to \Phi_{12}$, and correspondingly $v_{21} \to v_{12}$, $v_1 \to v_2$, where the new phase velocities are obtained from the expressions above by the substitution $\omega_1 \leftrightarrow \omega_2$ and $\bm\theta_{pk_1} \leftrightarrow \bm\theta_{pk_2}$.

In order to check the obtained numerical result we use a different identity, which can be obtained by the integration by parts
\begin{align}
	&\int_{-\infty}^{+\infty}d\left[ e^{i \Phi_{21}(\phi)}  \int_{-\infty}^{\phi}d\varphi \left(\frac{e \bm A(\varphi)}{m\xi}\right)^j  e^{i \Phi_1(\varphi)}\right]=0\,,\\
	&\int_{-\infty}^{+\infty}d\phi \left(\frac{e \bm A(\phi)}{m\xi}\right)^i e^{i \Phi_{21}(\phi)} \int_{-\infty}^{\phi}de^{i \Phi_1(\varphi)}=
	\int_{-\infty}^{+\infty}d\phi \left(\frac{e \bm A(\phi)}{m\xi}\right)^i e^{i \Phi_1(\phi)+i \Phi_{21}(\phi)}\nonumber
	\,.
\end{align}
The left-hand sides of these identities can be expressed in terms of the functions $f_{ij}$.

We use the same numerical method as in the previous subsection~\ref{sec:num1}. In particular, we use an analogous procedure for determining the number of points in the grid with a small modification. We use the maximum phase difference between $\Phi_{21}$ and ${\Phi_1}$. We also use \texttt{Compile}  in Wolfram Mathematica. On a standard personal computer, the computational time for $f_{ij}$ is $10^{-3}$ s for $N_0=10^3$. For $\omega\sim\varepsilon/100$ the accuracy is about $10^{-4}$--$10^{-3}$. Increasing the photon energy leads to a larger phase variation $\Delta\Phi$, which requires a larger number of grid points to maintain the same accuracy. For $\omega\sim\varepsilon$, the value $N_0=10^3$ may be insufficient, and $N_0$ must be increased to achieve the desired accuracy. The computational time for the differential probability $dW$~\eqref{eq:W} or $dW^{(1)}$~\eqref{eq:W1} for a single kinematic point is $10^{-2}$ s. Thus we can also use Monte Carlo methods for numerical integration over the photon angles. The analysis of different Monte Carlo methods implemented in Wolfram Mathematica shows that the best method for evaluating such peaked four-dimensional integrals over the photon angles is \texttt{QuasiMonteCarlo}. It allows one to obtain results with an accuracy of about $1\%$ using $10^6$ points.

We attach the \texttt{Wolfram Mathematica} notebook \texttt{Double Compton.nb} as an ancillary file. It contains the numerical implementation of the functions $f_{ij}$ and $g_{ij}$ as well as the differential probability $dW$ for the exact and soft photon results, Eqs.~\eqref{eq:W} and \eqref{eq:W1}.

\section{Conclusion}\label{sec:concl}
In this paper, we have applied the soft photon approximation to QED processes in a strong laser field. Building upon the formalism developed in our previous work \cite{Krachkov24}, we have considered nonlinear single Compton scattering, nonlinear double Compton scattering, and general $N$-photon emission, providing analytical expressions that simplify calculations in strong-field QED.

For single nonlinear Compton scattering, we have performed a detailed comparison between the soft photon approximation, the exact result, and the classical result. Our analysis shows that the soft photon approximation reproduces the exact spectrum with an accuracy of $\mathcal{O}(\omega/\varepsilon)$, while the classical approximation is valid only in a narrow region $\omega/\varepsilon \lesssim 0.01$ for the discussed kinematics. The difference between our approach and classical radiation theory is the exact treatment of the phase factor: while classical approaches neglect the quantum recoil effect, our approximation retains the recoil effects exactly in the phase, approximating only the preexponential factors. This distinction is physically important because the phase is integrated over the macroscopic laser pulse length, and even a small difference in the phase can change the final probability. The relative difference between the exact and soft photon spectra reaches approximately $25\%$ at $\omega/\varepsilon = 0.2$, showing that the method provides accurate results beyond the regime where classical theory fails. We note that the approximation is not valid at exactly zero emission angles, but this occurs only in a narrow angular region and does not affect the integrated spectrum.

For double nonlinear Compton scattering, we have derived analytical expressions for the differential probability in two kinematic regimes: when both photons are soft, and when one photon is soft while the other has arbitrary energy. Our formulas are expressed in terms of two-dimensional integrals. We have verified the accuracy of our results by comparing the two kinematic regimes, showing that the soft photon approximation maintains $\mathcal{O}(\omega/\varepsilon)$ accuracy even in the presence of interference effects. The two-dimensional integral is computed in a single pass over the grid by updating a primitive function incrementally, reducing the computational complexity from $O(N_{\text{grid}}^2)$ to $O(N_{\text{grid}})$, where $N_{\text{grid}}$ is the number of grid points.

We have also presented an analytical expression for the amplitude of $N$-photon emission within the soft photon approximation. Previous studies of multiphoton emission in a plane-wave background focused on the incoherent cascade contribution~\cite{PhysRevD.99.096018}, where the  probability factorizes into a product of independent single-photon emission probabilities. In contrast, the amplitude presented here accounts for the interference between different emission orderings. The amplitude is expressed as a sum over all permutations of emission orders, with each term given by an integral over a simplex. The simplex structure of the integrals and the factorization of the integrand allow one to compute the differential probability with $\mathcal{O}(N_{\text{grid}})$ complexity, similar to the double Compton case. This result provides a framework for studying cascade processes in strong-field QED, where an electron emits multiple photons. In the classical limit, the sum over permutations factorizes, reproducing the well-known independent emission of photons. The soft photon approximation breaks this factorization by capturing the quantum recoil effect through the modification of the electron momentum after each emission, which is reflected in the phase.

We have provided a numerical implementation in Wolfram Mathematica for single and double nonlinear Compton scattering. The code implements the general formulas derived in the paper and can be used for arbitrary collision geometries, laser pulse shapes, and electron energies with small modifications. The numerical complexity is reduced to $\mathcal{O}(N_{\text{grid}})$ using the factorization properties of the integrands, enabling efficient computation of differential probabilities. The code reproduces all numerical results presented in the paper and can be adapted to study other regimes of interest.

The method developed in this work is not restricted to the head-on ultra-relativistic regime. The general expressions derived in this paper are valid for arbitrary kinematics and can be applied to study nonlinear Compton scattering at large angles, different collision geometries, and arbitrary arbitrary plane-wave pulse shapes. The soft photon approximation can be applied for various QED processes involving additional soft photon emission. In particular, this approach is more suitable for calculation of the soft part of the virtual and real radiative corrections for the QED processes in the plane wave background.

\section*{Appendix}\label{sec:app} 
This appendix provides additional details on the formulas and notation used in the paper.

Let vector $\bm{n}$ define the propagation direction of the plane wave. Vector potential $A^{\mu}(\phi)$ depends only on  $\phi=t-\bm{n}\cdot\bm{x}.$ We introduce four four-dimensional quantities: $n^{\mu}=(1,\bm{n})$,
$\tilde{n}^{\mu}=(1,-\bm{n})/2$, and $a_{j}^{\mu}=(0,\bm{a}_{j})$, where $j=1,2$. The four-dimensional quantities $n^{\mu}$, $\tilde{n}^{\mu}$, and $a_{j}^{\mu}$
fulfill the completeness relation  
\begin{eqnarray*}
	\eta^{\mu\nu} & = & n^{\mu}\tilde{n}^{\nu}+\tilde{n}^{\mu}n^{\nu}-a_{1}^{\mu}a_{1}^{\nu}-a_{2}^{\mu}a_{2}^{\nu}\,.
\end{eqnarray*}
Note that $(n\tilde{n})=1$, $(n a_{j})=(\tilde{n}a_{j})=\tilde{n}^{2}=n^{2}=0$,
$a_{i}a_{j}=-\delta_{i,j}$. In what follows, we refer to the longitudinal ($n$) direction as the direction along $\bm{n}$ and to the transverse ($\perp$) plane as the plane spanned by the two perpendicular unit vectors $\bm{a}_{j}$. Coordinates:
\[
\phi=(nx)=t-x_{n}\,,\quad T=(\tilde{n}x)=(t+x_{n})/2\,,\quad\bm{x}_{\perp}^{i}=-(a_{i}x)\,,
\]

where $x_{n}=\bm{n}\cdot\bm{x}$. For an arbitrary four-vector $p$, we
introduce $p_{-}=(np)=p^{0}-p_{n}$, $p_{+}=(\tilde{n}p)=(p^{0}+p_{n})/2\,$,
and $\bm{p}_{\perp}=(p_{\perp,1},p_{\perp,2})=-((pa_{1}),(pa_{2}))=(\bm{p}\cdot\bm{a}_{1},\bm{p}\cdot\bm{a}_{2})$.
Thus, 
\[
px=p_{\mu}x_{\nu}\eta^{\mu\nu}=(xn)(p\tilde{n})+(pn)(x\tilde{n})-\bm{x}_{\perp}\cdot\bm{p}_{\perp}=p_{-}T+p_{+}\phi-\bm{x}_{\perp}\cdot\bm{p}_{\perp}\,.
\]

The momenta operators on this basis have the form
\[
\begin{array}{c}
P_{\phi}=-i\partial_{\phi}=-(\tilde{n}P)=-(i\partial_{t}-i\partial_{x_{n}})/2\quad P_{T}=-i\partial_{T}=-(nP)=-(i\partial_{t}+i\partial_{x_{n}}),\\
\bm{P}_{\perp}=(P_{\perp,1},P_{\perp,2})=-i(\bm{a}_{1}\cdot\bm{\nabla},\bm{a}_{2}\cdot\bm{\nabla})\,.
\end{array}
\]
They satisfy the following commutation relation:
\[
[\phi,P_{\phi}]=[T,P_{T}]=i\,,\quad[X_{\perp,j},P_{\perp,k}]=i\delta_{jk}\,,
\]

which are equivalent to the commutation relations $[X^{\mu},P^{\nu}]=-i\eta^{\mu\nu}$,
with $P^{\mu}=i\partial^{\mu}$.

The Volkov states $U_{p}(x)$ and $V_{p}(x)$ can be written as
\begin{align}\label{eq:vawe_func}
	U_{p}(x)&=\bigg[1+\frac{e\hat{n}\hat{A}(\phi)}{2p_{-}}\bigg]u_{p}\text{e}^{-i\int_{0}^{x}(\pi^\mu_p(x')+eA^\mu(x'))dx'_\mu }\,,\nonumber\\
	\bar{U}_{p}(x)&=\bar{u}{}_{p}\bigg[1-\frac{e\hat{n}\hat{A}(\phi)}{2p_{-}}\bigg]\text{e}^{i\int_{0}^{x}(\pi^\mu_p(x')+eA^\mu(x'))dx'_\mu }\,,\nonumber\\
	V_{p}(x)&=\bigg[1-\frac{e\hat{n}\hat{A}(\phi)}{2p_{-}}\bigg]v_{p}\text{e}^{-i\int_{0}^{x}(\pi^\mu_{-p}(x')+eA^\mu(x'))dx'_\mu }\,,\nonumber\\
	\bar{V}_{p}(x)&=\bar{v}{}_{p}\bigg[1+\frac{e\hat{n}\hat{A}(\phi)}{2p_{-}}\bigg]\text{e}^{i\int_{0}^{x}(\pi^\mu_{-p}(x')+eA^\mu(x'))dx'_\mu }\,,
\end{align}
where $u_{p}$ and $v_{p}$ are the free spinors. Here we have introduced the notation $\hat{l}=\gamma^{\mu}l_{\mu}$
for a generic four-vector $l^{\mu}$, with $\gamma^{\mu}$ being the Dirac matrices.

We note that the choice of the lower integration limit $0$ in Eq.~\eqref{eq:vawe_func} differs from the standard convention where the integral is taken from $-\infty$  (see, e.g., Ref.~\cite{LL4}). This choice introduces an additional constant phase factor in each Volkov state, which depends on the initial conditions at $x=0$. However, this constant phase cancels in all physical observables, since it appears as an overall phase factor in the matrix element and disappears upon taking the absolute square $|S|^2$. 

 The classical kinetic four-momentum of an electron in the plane wave $A^{\mu}(\phi)$ has the form
\begin{equation}
\pi_{p}^{\lambda}(\phi)=p^{\lambda}-eA^{\lambda}(\phi)+\frac{e(pA(\phi))}{p_{-}}n^{\lambda}-\frac{e^{2}A^{2}(\phi)}{2p_{-}}n^{\lambda}\,,
\end{equation}
with $\lim_{\phi\rightarrow\pm\infty}\pi_{p}^{\lambda}(\phi)=p^{\lambda}$.
Note that $\pi_{p}(\phi)^{2}=p^{2}$ and $\pi_{p-}(\phi)=p_{-}$. Note that the quantity $\pi_{-p}(\varphi) $ that occurs in  spinors $V$  can be represented as $-\pi_{p}(\varphi)$ with the substitution $e\rightarrow-e$, where $e$ is an electron charge.

\bibliography{sphoton2}
\end{document}